\documentclass[fleqn,usenatbib]{mnras}

\usepackage{mathptmx}
\usepackage[T1]{fontenc}

\DeclareRobustCommand{\VAN}[3]{#2}
\let\VANthebibliography\thebibliography
\def\thebibliography{\DeclareRobustCommand{\VAN}[3]{##3}\VANthebibliography}

\usepackage{graphicx}	% Including figure files
\usepackage{amsmath}	% Advanced maths commands
\usepackage{amssymb}	% Extra maths symbols
\usepackage[normalem]{ulem} % for \sout
\usepackage{xcolor}
\newcommand{\boldvec}[1]{\vec{\mbox{\boldmath{$#1$}}}}% Katia
\newcommand{\boldhat}[1]{\hat{\mbox{\boldmath{$#1$}}}}% Katia
\title[The correct sense of Faraday rotation]{The correct sense of Faraday
rotation}

\author[Ferri\`{e}re et al.]{
K. Ferri\`{e}re,$^{1}$\thanks{E-mail: katia.ferriere@irap.omp.eu}
J. L. West,$^{2}$
T.~R. Jaffe$^{3}$
\\
$^{1}$Institut de Recherche en Astrophysique et Plan\'{e}tologie (IRAP),
Universit\'{e} de Toulouse, CNRS, 9 avenue du Colonel Roche, BP 44346,
31028 Toulouse Cedex 4, France\\
$^{2}$Dunlap Institute for Astronomy and Astrophysics University of Toronto,
Toronto, ON M5S 3H4, Canada\\
$^{3}$NASA Goddard Space Flight Center, Greenbelt, MD 20771, USA
}

\date{Accepted XXX. Received YYY; in original form ZZZ}

\pubyear{2015}

\begin{document}
\label{firstpage}
\pagerange{\pageref{firstpage}--\pageref{lastpage}}
\maketitle

\begin{abstract}
The phenomenon of Faraday rotation of linearly polarized synchrotron emission in a
magneto-ionized medium has been understood and studied for decades.  But since the
sense of the rotation itself is irrelevant  in most contexts, some uncertainty and
inconsistencies have arisen in the literature about this detail.  Here, we start
from basic plasma theory to describe the propagation of polarized emission from a
background radio source through a magnetized, ionized medium in order to rederive
the correct sense of Faraday rotation.  We present simple graphics to illustrate
the decomposition of a linearly polarized wave into right and left circularly
polarized modes, the temporal and spatial propagation of the phases of those
modes, and the resulting physical rotation of the polarization orientation.  We
then re-examine the case of a medium that both Faraday-rotates and emits polarized
radiation and show how a helical magnetic field can construct or destruct the
Faraday rotation.  This paper aims to resolve a source of confusion that has
arisen between the plasma physics and radio astronomy communities and to help
avoid common pitfalls when working with this unintuitive phenomenon.
\end{abstract}

\begin{keywords}
ISM -- magnetic fields -- polarization
\end{keywords}

%%%%%%%%%%%%%%%%%%%%%%%%%%%%%%%%%%%%%%%%%%%%%%%%%%

%%%%%%%%%%%%%%%%% BODY OF PAPER %%%%%%%%%%%%%%%%%%

\section{Introduction}
Faraday rotation of linearly polarized synchrotron radiation at radio wavelengths
is one of the primary tools used to study galactic and extragalactic magnetic
fields. The Faraday rotation measure (RM) is commonly derived from observational
data by taking measurements at a variety of wavelengths, $\lambda$, and examining
how the polarization angle (PA) changes. In the simplest case of a background
polarized source, a plot of PA vs $\lambda^2$ shows a linear relation and the
slope of the line gives the RM. 
The conventions now widely known and adopted in observational radio astronomy were
first put forth by \cite{manchester_72}: when ${\rm RM} > 0$, the magnetic field,
$\boldvec{B}$, is on average directed toward the observer, and when ${\rm RM} <
0$, $\boldvec{B}$ is on average directed away from the observer.
However, the physical sense of rotation of the electric field vector of the wave
is not so well known, as this information is generally not relevant in
astrophysical problems. Many astrophysics textbooks contain a comparatively short
section on this topic and tend to gloss over some of the finer points that are
critical to a complete understanding.
Even the astrophysics textbooks that do provide a detailed derivation
\citep{spitzer_78, rybicki&l_79, bowers&d_84, shu_91, elitzur_92} tend to lay the
emphasis on the physical mechanism rather than the actual sense of rotation.

A complete mathematical derivation of Faraday rotation can be found in several
textbooks of plasma physics \citep[e.g.,][]{nicholson_83, chen_16} or optics and
radiation theory \citep[e.g.,][]{stone_63, papas_65, stutzman_93, goldstein_11,
collett&s_12}. However, few discuss the actual sense of rotation explicitly, and
those who do could easily leave radio astronomers confused, either because they
measure the sense of rotation from a different point of view \citep[][who use the
same conventions as optical astronomers; see
Sect.~\ref{sec:confusion_modes}]{stone_63, papas_65, collett&s_12} or because
their reasoning is flawed \citep[][see
Sect.~\ref{sec:confusion_examples}]{chen_16}.

There are cases where knowing the sense of Faraday rotation is of critical
importance. For instance, several studies have looked at a possible link between
Faraday rotation and magnetic helicity \citep{volegova&s_10,brandenburg&s_14}. 
Our own recent work \citep{west&hfw_20} confirmed the existence of such a link,
but the correlation that we measured systematically had the opposite sign
to that obtained in certain previous studies.
In fact, we were able to recover the results of these previous studies
provided we flipped the sense of Faraday rotation.
This naturally led us to question our adopted sense of Faraday rotation.
We consulted several reference textbooks (including those cited above) and we
asked several leading experts, both in plasma physics and in radio astronomy.
We were quite surprised by the variety, and sometimes the uncertainty, in the
answers
we found or received.
Thus, in the course of our "investigation", we realized that
what was supposed to be a trivial question
did not have a simple and unanimous answer in the community.
The purpose of this paper is to clarify the situation and, whenever possible,
either reconcile the different approaches or explain why and where they disagree.

In Sect.~\ref{sec:mathderivation}, we rederive the correct sense of Faraday
rotation
as well as the exact expression of the Faraday RM
from basic plasma theory.
Along the way, we provide detailed physical interpretations and simple graphical
illustrations of our equations, while the more mathematical aspects of the
derivation are relegated to Appendices~\ref{sec:dispersion_relation} and
\ref{sec:Efield_expression}.
In Sect.~\ref{sec:confusion}, we discuss four possible sources of confusion
and explain how they can lead to erroneous reasoning and/or faulty conclusions.
In Sect.~\ref{sec:helicity}, we return to the problem that motivated the present
paper
and re-examine the link between Faraday rotation and magnetic helicity.
In Sect.~\ref{sec:conclusions}, we conclude our study.

Throughout the paper, we use the IAU reference frame for polarization,\footnote{
IAU General Assembly Meeting, 1973, Commission 40 (Radio Astronomy), 8.
POLARIZATION DEFINITIONS.
%%''A working Group chaired by Westerhout was convened to discuss the definition of polarization
%%brightness temperatures used in the description of polarized extended objects and the galactic
%%background. The following resolution was adopted by Commissions 25 and 40: 'RESOLVED,
%%that the frame of reference for the Stokes parameters is that of Right Ascension and Declination
%%with the position angle of electric-vector maximum, $\theta$, starting from North and increasing through
%%East. Elliptical polarization is defined in conformity with the definitions of the Institute of Electrical
%%and Electronics Engineers (IEEE Standard 211, 1969). This means that the polarization of incoming
%%radiation, for which the position angle, $\theta$, of the electric vector, measured at a fixed point in space,
%%increases with time, is described as right-handed and positive.''
}
which is shown in Figure 1.
Accordingly, we take the position angle of linear polarization, simply referred to
as the polarization angle, $\psi$, to increase counterclockwise in the sky,
starting from North.
As a note of caution for CMB radio astronomers, let us mention that the IAU
convention for the sign of $\psi$ is opposite to that chosen for cosmological
analysis in much of the CMB community.

\begin{figure}
\centering
\includegraphics[width=0.45\textwidth]{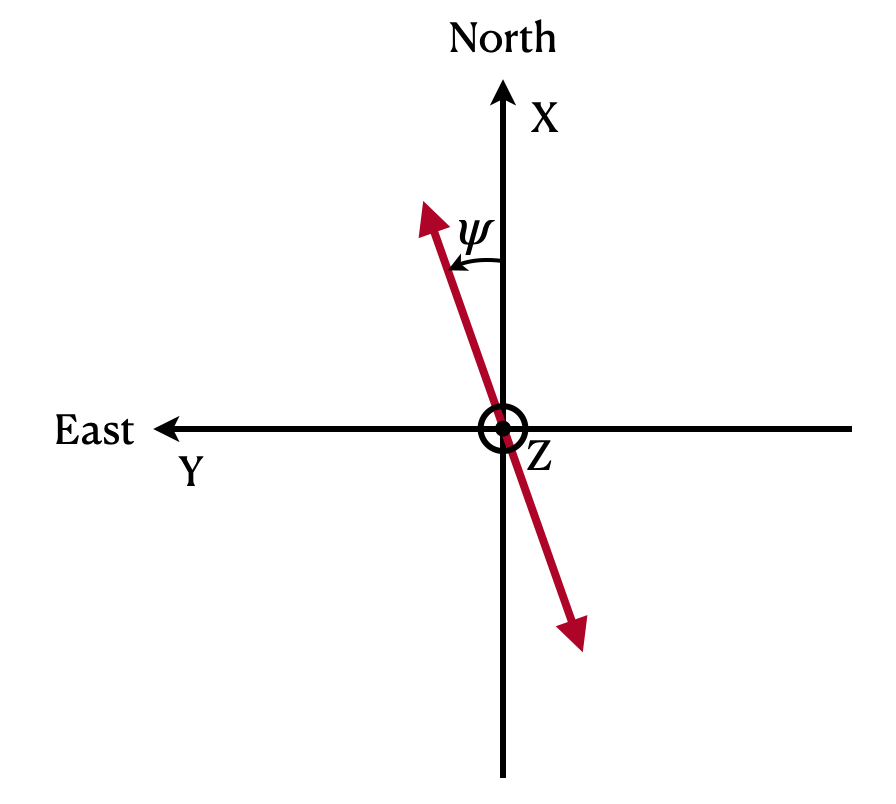}
\caption{
IAU coordinate and polarization conventions used in this paper.
$X$ and $Y$ are the Cartesian coordinates in the plane of the sky
(increasing toward North and East, respectively),
$Z$ is the line-of-sight coordinate (increasing toward the observer),
the red double-headed arrow indicates the orientation of linear polarization,
and $\psi$ denotes the associated position angle, also referred to as the
polarization angle (measured counterclockwise from North).
}
\label{fig:IAU}
\end{figure}

\section{Mathematical derivation}
\label{sec:mathderivation}

\subsection{The two circularly polarized modes of parallel propagation}
\label{sec:mathderivation_circular}

Consider a cold, magnetized plasma with electron density $n_{\rm e}$
and magnetic field $\boldvec{B}$. %%$\boldvec{B} = B \, \boldvec{e}_z$.
Two characteristic frequencies of this plasma are
the plasma frequency, $\omega_{\rm e} =
\sqrt{\frac{\displaystyle 4\pi n_{\rm e} e^2}{\displaystyle m_{\rm e}}}$,
and the electron gyro-frequency, 
$\Omega_{\rm e} = \frac{\displaystyle q_{\rm e} B}{\displaystyle m_{\rm e} c}$,
where $m_{\rm e}$ is the mass of the electron,
$q_{\rm e} = -e$ the electric charge of the electron ($q_{\rm e} < 0$),
$B$ the magnetic field strength ($B>0$), and $c$ the speed of light.
Note that $\Omega_{\rm e}$, as defined by plasma physicists, is negative. However,
for convenience, $\Omega_{\rm e}$ is often redefined as a positive quantity
($\Omega_{\rm e} = \frac{\displaystyle e B}{\displaystyle m_{\rm e} c}$) outside
the plasma community.
Here, to avoid any possible confusion, we will work with the absolute value of
$\Omega_{\rm e}$: $|\Omega_{\rm e}| = \frac{\displaystyle e B}{\displaystyle
m_{\rm e} c}$, which is unambiguously positive.
In the interstellar medium (ISM),
most of the free electrons reside in the warm ionized medium (WIM),
where $n_{\rm e} \sim 0.2~{\rm cm}^{-3}$ and $B \sim 5~\mu{\rm G}$,
leading to $\omega_{\rm e} \sim 25~{\rm kHz}$ and $|\Omega_{\rm e}| \sim 90~{\rm
Hz}$ \citep[e.g.,][]{ferriere_20}.

Let us now study the propagation of a radio electromagnetic wave
with given angular frequency $\omega > 0$ (imposed at the source),
corresponding to frequency $\nu = \frac{\displaystyle \omega}{\displaystyle
2\pi}$.
Assume that this wave has wave vector $\boldvec{k}$ (directed from the source to
the observer), 
corresponding to wavenumber $k = |\boldvec{k}| > 0$,
propagation direction $\boldhat{e}_k = \frac{\displaystyle
\boldvec{k}}{\displaystyle k}$,
and wavelength $\lambda = \frac{\displaystyle 2\pi}{\displaystyle k}$.
The wavenumber in any propagation direction is not imposed at the source, but it
is given by the dispersion relation, which depends on the properties of the
traversed medium.

Let us first focus on the case of parallel propagation, when $\boldvec{k}
\parallel \boldvec{B}$.
The dispersion relation can then be written as (see
Appendix~\ref{sec:dispersion_relation})
\begin{equation}
\omega^2 \ = \ c^2 \, k^2
\ + \ \frac{\omega_{\rm e}^2}
{1 \mp \frac{\displaystyle |\Omega_{\rm e}|}{\displaystyle\omega}} \ ,
\label{eq_DR}
\end{equation}
where the $\mp$ sign in the denominator arises from the existence of two
solutions:
a right circularly polarized mode,
for which the electric field vector of the wave rotates
in a right-handed sense about $\boldvec{B}$ (upper sign),
and a left circularly polarized mode,
for which the electric field vector of the wave rotates
in a left-handed sense about $\boldvec{B}$ (lower sign).
Potential confusion regarding the concepts and definitions of right and left
circularly polarized modes will be discussed in Sect.~\ref{sec:confusion_modes}.

The waves of interest here typically have $\nu \gtrsim 100~{\rm MHz}$,
so that $\omega_{\rm e}, \, |\Omega_{\rm e}| \lll \omega$.
Under these conditions, Eq.~(\ref{eq_DR}) can be approximated by
\begin{equation}
\omega^2
\ = \ c^2 \, k^2 + \omega_{\rm e}^2
\pm \frac{\omega_{\rm e}^2 \, |\Omega_{\rm e}|}{\omega} \ \cdot
\label{eq_DR_2nd}
\end{equation}
The first term in the right-hand side represents electromagnetic wave propagation
in free space.
The next two terms describe the diamagnetic effect of the plasma,
with a dominant contribution to the electric current arising from the electric
force (second term)
and a  much weaker contribution arising from the magnetic force (third term).
The latter adds to [subtracts from] the electric force
when the electric field vector rotates in the same sense as [the opposite sense
to]
the natural gyration of electrons about $\boldvec{B}$.

The expression of the phase velocity, $V_\phi$, directly follows from
Eq.~(\ref{eq_DR_2nd}):
\begin{equation}
V_\phi
\ \equiv \ \frac{\omega}{k}
\ = \ c \
\left( 1 +\frac{\omega_{\rm e}^2}{2 \, \omega^2}
\pm \frac{\omega_{\rm e}^2 \, |\Omega_{\rm e}|}{2 \, \omega^3}
\right) \ ,
\label{eq_phase_vel}
\end{equation}
where again the upper [lower] sign in the third term pertains to the right (R)
[left (L)] mode.
%%Clearly, the right (R) and left (L) modes have slightly different phase velocities,
Clearly, although both modes have the same angular frequency $\omega$ (imposed at
the source),
they have slightly different phase velocities,
with $V_{\phi,{\rm R}} > V_{\phi,{\rm L}}$.
As a result, if they start off with the same phase at the source, a phase
difference will arise and grow between them as they propagate away from the
source.
The sign of this phase difference, which is critical to the sense of Faraday
rotation,
depends on how exactly the phase is defined.

In a homogeneous plasma, the phase can be defined as
either $\phi = (\omega\,t - k\,s + \varphi_0)$ or $\phi = (k\,s - \omega\,t +
\varphi_0)$,
where $t$ denotes the time since a certain initial time,
$s$ denotes the distance from the radio source along the propagation direction,
$\boldhat{e}_k$,
and $\varphi_0$ is an arbitrary phase offset.
In an inhomogeneous plasma, the above expressions should be replaced by
$\phi = \left( \omega\,t - \int_0^s k \, ds' + \varphi_0 \right)$
and $\phi = \left( \int_0^s k \, ds' - \omega\,t + \varphi_0 \right)$,
respectively.
For convenience, we set the phase offset to $\varphi_0 = 0$,
such that $\phi=0$ at the source ($s=0$) and initial time ($t=0$).

\begin{table}
\begin{tabular}{llcc}
\hline\hline
\noalign{\medskip}
\multicolumn{2}{c}{Definition of the phase} & 
Behavior & Behavior \\
& & with increasing $t$ & with increasing $s$ \\
& & at constant $s$ & at constant $t$ \\
\noalign{\medskip}
\hline
\noalign{\medskip}
$\displaystyle \phi = \omega\,t - \int_0^s k \, ds'$ & (Eq.~\ref{eq_phase}) & 
$\phi$ increases & $\phi$ decreases \\
\noalign{\medskip}
$\displaystyle \phi = \int_0^s k \, ds' - \omega\,t$ & (Eq.~\ref{eq_phase_bis}) & 
$\phi$ decreases & $\phi$ increases \\
\noalign{\medskip}
\hline\hline
\end{tabular}
\caption{
Two alternative definitions of the phase, $\phi$, of a radio wave
with angular frequency $\omega$ and wavenumber $k$,
at time $t$ and distance $s$ from the source (first column),
and corresponding behaviors with increasing $t$ at constant $s$ (second column)
and with increasing $s$ at constant $t$ (third column).
}
\label{table_phase}
\end{table}

With the first definition of the phase,
$\phi$ at a given distance $s$ increases with increasing time $t$,
while $\phi$ at a given time $t$ decreases with increasing distance $s$
(see Table~\ref{table_phase});
this means that $\phi$ can be taken equal to the angle 
through which the electric field vector has rotated
(in a right-handed [left-handed] sense about $\boldvec{B}$ for the right [left]
mode)
from its initial direction at the source.
With the second definition,
$\phi$ at a given distance $s$ decreases with increasing time $t$,
while $\phi$ at a given time $t$ increases with increasing distance $s$;
this means that $\phi$ can be taken equal to {\it minus}
the angle through which the electric field vector has rotated
from its initial direction at the source.
An alternative interpretation of $\phi$ in the latter case will be provided in
Sect.~\ref{sec:confusion_phase}.

Here, and for the rest of the paper (except in Sect.~\ref{sec:confusion_phase}),
we adopt the first definition, which directly gives the angle of the electric
field vector and which conforms to the IEEE standard:\footnote{
IEEE Standard Definitions of Terms for Radio Wave Propagation (IEEE Std 211-1997).
}
\begin{equation}
\phi \ = \ \omega\,t - \int_0^s k \, ds' \ \cdot
\label{eq_phase}
\end{equation}
However, we emphasize that both definitions are valid
and both lead to the same sense of Faraday rotation.
This will become clearer in Sect.~\ref{sec:confusion_phase},
where we discuss the implications of adopting the opposite definition of the
phase.

\subsection{Propagation of a linearly polarized wave}
\label{sec:mathderivation_linear}

Consider a source of linearly polarized synchrotron radiation, 
e.g., a Galactic pulsar, at a given angular frequency $\omega$.
In reality, synchrotron radiation is not fully polarized, but we will only retain
its linearly polarized component, which in practice is measured through the Stokes
parameters $Q$ and $U$.\footnote{
In general, the polarization state of electromagnetic radiation can be described
by the four Stokes parameters: the total intensity, $I$; the linear polarization
parameters, $Q$ and $U$, whose quadratic sum gives the linearly polarized
intensity, $|P|$ (see Eq.~\ref{eq_synchr_PI} below), and whose ratio yields the
polarization angle, $\psi$ (see Eq.~\ref{eq_synchr_polangle}); and the circular
polarization parameter, $V$, which gives the circularly polarized intensity.
Synchrotron radiation is partially linearly polarized, with $|P| < I$ and $V=0$.
}
Accordingly, we consider that the electric field vector, $\boldvec{E}_\ell$,
oscillates at angular frequency $\omega$ along an axis with given orientation,
which defines the polarization orientation.
For future reference, we select one of the two unit vectors in the polarization
orientation at the source (subscript $\star$) and denote it by
$\boldhat{e}_\star$.
If we choose the initial time appropriately, we can then write the electric field
vector at the source as
\begin{equation}
\boldvec{E}_{\ell \star} = E_0 \ \cos (\omega\,t) \ \boldhat{e}_\star \ \cdot
\label{eq_Efield_lin_source}
\end{equation}

A linearly polarized wave with angular frequency $\omega$ propagating parallel to
the magnetic field, $\boldvec{B}$, can be seen as the superposition of a right (R)
and a left (L) circularly polarized mode with the same $\omega$.
The electric field vectors of the right and left modes, $\boldvec{E}_{\rm R}$ and
$\boldvec{E}_{\rm L}$, rotate in opposite senses, symmetrically with respect to
the polarization orientation, in such a way that their vector sum equals the
electric field vector of the linearly polarized wave, $\boldvec{E}_\ell$:
\begin{equation}
\boldvec{E}_\ell = \boldvec{E}_{\rm R} + \boldvec{E}_{\rm L} \ \cdot
\label{eq_Efield_lin_sum}
%%\label{eq_electricvec}
\end{equation}
At the source and initial time, $\boldvec{E}_\ell$ is at its maximum extent in the
$+\boldhat{e}_\star$ direction (see Eq.~\ref{eq_Efield_lin_source}), which implies
that $\boldvec{E}_{\rm R}$ and $\boldvec{E}_{\rm L}$ are both pointing in the
$+\boldhat{e}_\star$ direction.
Therefore, $\boldhat{e}_\star$ represents the initial direction of
$\boldvec{E}_{\rm R}$ [$\boldvec{E}_{\rm L}$] at the source, which, as explained
above Eq.~(\ref{eq_phase}), defines the reference direction from which the current
angle of $\boldvec{E}_{\rm R}$ [$\boldvec{E}_{\rm L}$] (measured in a right-handed
[left-handed] sense about $\boldvec{B}$) can be equated to the current phase of
the right [left] mode, $\phi_{\rm R}$ [$\phi_{\rm L}$], as defined by
Eq.~(\ref{eq_phase}).

At the source ($s=0$), the right and left modes have the same phase,
$\phi_\star(t) = \omega\,t$.
At a distance $s$ from the source, the two modes have a phase difference,
$\Delta \phi \equiv \phi_{\rm R} - \phi_{\rm L}$, given by
\begin{eqnarray}
\Delta \phi
& = & - \int_0^s \Delta k \ ds'
\nonumber \\
& = & \int_0^s \frac{\Delta V_\phi}{V_\phi} \ k \ ds'
\nonumber \\
& = & \int_0^s \frac{\omega_{\rm e}^2 \, |\Omega_{\rm e}|}{\omega^3} \ k \ ds' \ ,
\label{eq_phase_diff}
\end{eqnarray}
where we have successively used Eq.~(\ref{eq_phase}) with $\Delta \omega \equiv
\omega_{\rm R} - \omega_{\rm L} = 0$ (remember that both modes have the same
$\omega$) and $\Delta k \equiv k_{\rm R} - k_{\rm L}$, the first equality in
Eq.~(\ref{eq_phase_vel}) with $\Delta V_\phi \equiv V_{\phi,{\rm R}} -
V_{\phi,{\rm L}}$, and the second equality in Eq.~(\ref{eq_phase_vel}) with the
upper [lower] sign in the last term corresponding to $V_{\phi,{\rm R}}$
[$V_{\phi,{\rm L}}$].
The phase difference between the right and left modes is positive, which means
that, at any distance $s$ from the source,
the right mode is more advanced in phase than the left mode.

\begin{figure*}
\begin{minipage}{\textwidth}
\centering
\includegraphics[width=\textwidth]{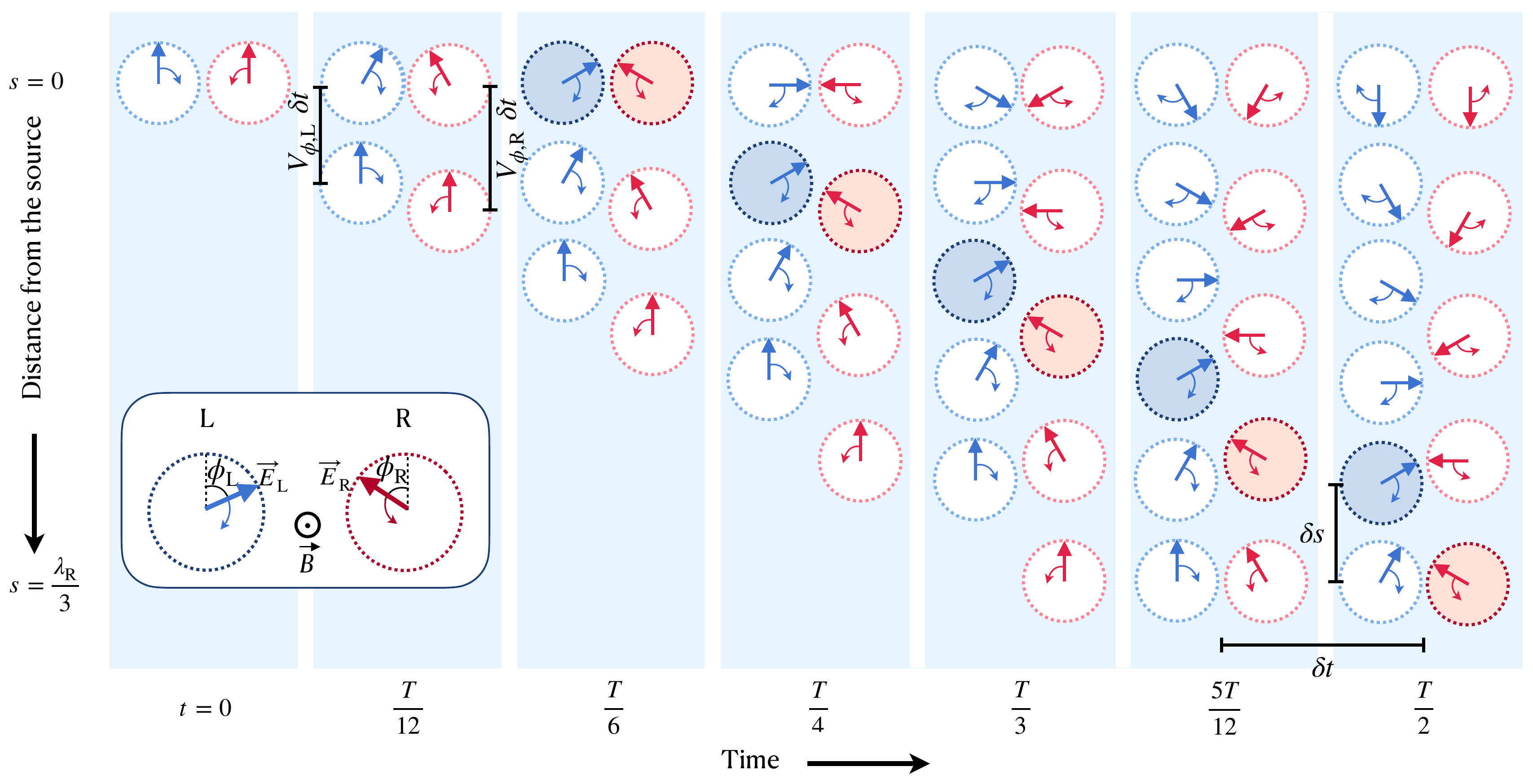}
\caption{
Cartoon illustrating the propagation of the phases $\phi_{\rm L}$ and $\phi_{\rm
R}$ of the left (blue) and right (red)  circularly polarized components of a
linearly polarized radio wave with period $T = \frac{\displaystyle
2\pi}{\displaystyle \omega}$.
Time, $t$, increases rightward, by steps of $\frac{T}{12}$, from the initial time
($t=0$) to half a period ($\frac{T}{2}$).
Distance from the source, $s$, increases continuously downward, 
from the source ($s=0$) to a third of the right mode's wavelength ($s =
\frac{\lambda_{\rm R}}{3}$).
With the phase of a mode defined through Eq.~(\ref{eq_phase}),$^{\rm a}$
$\phi_{\rm L}$ and $\phi_{\rm R}$ can be identified with the angles through which
the electric field vectors of the left and right modes, $\boldvec{E}_{\rm L}$ and
$\boldvec{E}_{\rm R}$, have respectively rotated (in a left-handed and
right-handed sense about the magnetic field, $\boldvec{B}$) from their common
initial direction at the source
(see inset in the lower-left corner).
The planes in which $\boldvec{E}_{\rm L}$ and $\boldvec{E}_{\rm R}$ rotate
are actually perpendicular to the propagation direction (the $s$-axis) and hence
perpendicular to the plane of the page.
However, for the reader to see them face-on, we inclined them by $90^\circ$, as if
$\boldvec{B}$ were pointing toward the reader.
The inclined views show $\boldvec{E}_{\rm L}$ (thick, blue arrows) and
$\boldvec{E}_{\rm R}$ (thick, red arrows), together with their respective sense of
rotation (thin, blue and red curved arrows) and the circles traced out by their
rotation (dotted, blue and red circles).
At the source (top row), $\boldvec{E}_{\rm L}$ and $\boldvec{E}_{\rm R}$ rotate
symmetrically with respect to their common initial direction (far-left column),
consistent with $\phi_{\rm L} = \phi_{\rm R}$.
Each $(\boldvec{E}_{\rm L},\boldvec{E}_{\rm R})$ pair plotted at the source can be
followed step by step as $\boldvec{E}_{\rm L}$ and $\boldvec{E}_{\rm R}$ propagate
away from the source at the phase velocities of the left and right modes,
$V_{\phi,{\rm L}}$ and $V_{\phi,{\rm R}}$, respectively.
For instance, the propagation of the $(\boldvec{E}_{\rm L},\boldvec{E}_{\rm R})$
pair starting at the source at time $\frac{T}{6}$, with $\phi_{\rm L} = \phi_{\rm
R} = \frac{\pi}{3}$, can be followed through the highlighted sequence.
Since $V_{\phi,{\rm R}} > V_{\phi,{\rm L}}$,$^{\rm b}$ $\boldvec{E}_{\rm R}$
propagates faster than $\boldvec{E}_{\rm L}$, and the right mode is ahead in phase
relative to the left mode.
$\delta s$ and $\delta t$ represent the spatial and temporal phase lags of the
left mode with respect to the right mode at $t = \frac{T}{2}$ and $s =
\frac{\lambda_{\rm R}}{3}$.
Figure produced with the help of Theophilus Britt Griswold (NASA Goddard).
}
\label{fig:phase_propa}
\parbox{\textwidth}
{
\begin{flushleft}
\scriptsize{
$^{\rm a}\ $ If the phase of a mode was defined through Eq.~(\ref{eq_phase_bis}),
the figure would look exactly the same, with this difference that the angles of
the electric field vectors (which themselves would remain unchanged) would be
labeled $-\phi_{\rm L}$ and $-\phi_{\rm R}$ instead of $\phi_{\rm L}$ and
$\phi_{\rm R}$.
\\
$^{\rm b}\ $ Here, for illustrative purposes, we took $V_{\phi,{\rm R}} = 1.25 \
V_{\phi,{\rm L}}$. In reality, the phase velocity difference between the right and
left modes is tremendously smaller, with $\frac{\displaystyle \Delta
V_\phi}{\displaystyle V_\phi} = \frac{\displaystyle \omega_{\rm e}^2 \,
|\Omega_{\rm e}|}{\displaystyle \omega^3} \lll 1$ (see Eq.~\ref{eq_phase_vel}).
For instance, with $\omega = 1~{\rm GHz}$, $\omega_{\rm e} = 25~{\rm kHz}$, and
$|\Omega_{\rm e}| = 90~{\rm Hz}$ (see beginning of
Sect.~\ref{sec:mathderivation_circular}), $\frac{\displaystyle \omega_{\rm e}^2 \,
|\Omega_{\rm e}|}{\displaystyle \omega^3} \simeq 2 \times 10^{-19}$.
}
\end{flushleft}
}
\end{minipage}
\end{figure*}

This can be understood physically with the help of Figure~\ref{fig:phase_propa},
which shows how the right and left modes propagate their phases, $\phi_{\rm R}$
and $\phi_{\rm L}$ (equal to the angles through which the electric field vectors
$\boldvec{E}_{\rm R}$ and $\boldvec{E}_{\rm L}$ have respectively rotated from
their common initial direction at the source, $\boldhat{e}_\star$), away from the
source.
At the source, both modes at any time $t_\star$ have the same phase,
$\phi_\star(t_\star) = \omega\,t_\star$, which is an increasing function of
$t_\star$. 
Both modes propagate this phase $\phi_\star(t_\star)$ away from the source at
their own phase velocities.
Since the right mode has a slightly larger phase velocity,
it is faster to propagate $\phi_\star(t_\star)$ out to a given distance $s$ from
the source.
In other words, a given $\phi_\star(t_\star)$ reaches $s$
earlier for the right mode than for the left mode.
By the time $t$ that $\phi_\star(t_\star)$ reaches $s$ for the right mode,
$\phi_\star(t_\star)$ is behind, say, at $s - \delta s$, for the left mode
and the phase of the left mode at $s$ is behind $\phi_\star(t_\star)$, say,
equal to $\phi_\star(t_\star - \delta t)$.
Mathematically, $\phi_\star(t_\star) = \phi_{\rm R}(t,s) = \phi_{\rm L}(t,s -
\delta s)$ and $\phi_{\rm L}(t,s) = \phi_\star(t_\star - \delta t) = \phi_{\rm
R}(t - \delta t,s)$.
In Figure~\ref{fig:phase_propa}, the propagation of $\phi_\star(t_\star)$ is
illustrated for $t_\star = \frac{T}{6}$ by the highlighted sequence of electric
field vectors at phase $\phi_\star(t_\star) = \frac{\pi}{3}$. 
The spatial lag, $\delta s$, and time lag, $\delta t$, are shown for $t =
\frac{T}{2}$ and $s = \frac{\lambda_{\rm R}}{3}$,
with $\lambda_{\rm R}$ the wavelength of the right mode.

\begin{figure*}
\centering
\includegraphics[width=\textwidth]{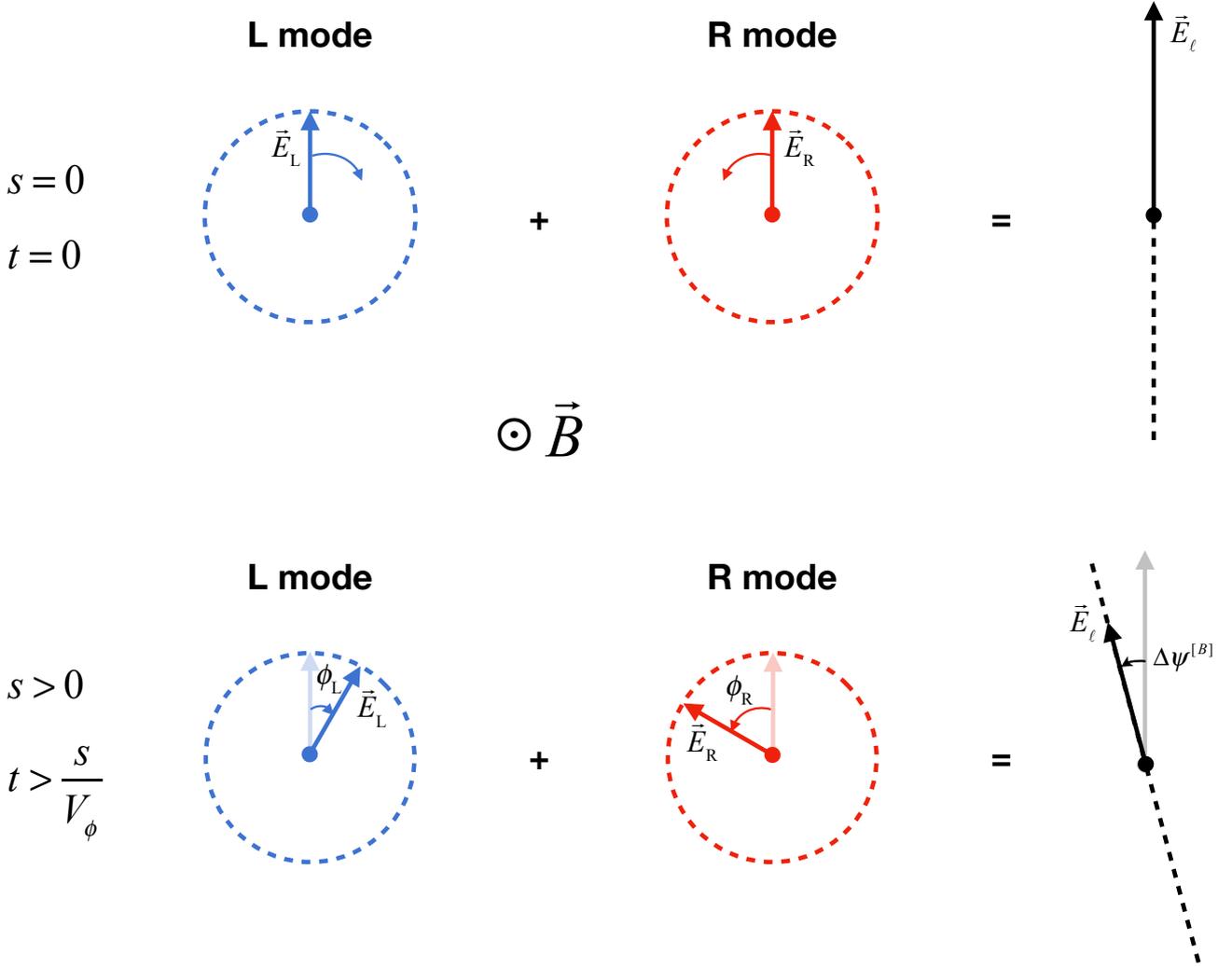}
\caption{ 
Cartoon showing the electric field vectors of a linearly polarized radio wave,
$\boldvec{E}_\ell$ (far right) and of its left (blue) and right (red) circularly
polarized components, $\boldvec{E}_{\rm L}$ and $\boldvec{E}_{\rm R}$, as defined
with respect to the magnetic field, $\boldvec{B}$ (plasma physics convention), and
as seen when $\boldvec{B}$ points toward the reader (as in
Figure~\ref{fig:phase_propa}).
The wave vector, $\boldvec{k}$, is parallel to $\boldvec{B}$, but its actual
direction ($+\boldvec{B}$ or $-\boldvec{B}$) is irrelevant.
Top row: at the source ($s=0$) and initial time ($t=0$).
Bottom row: at distance $s$ from the source and time $t$ greater than the phase
travel time to $s$.
The top and bottom rows correspond to the $(\boldvec{E}_{\rm L},\boldvec{E}_{\rm
R})$ pairs plotted in the upper-left and lower-right corners of
Figure~\ref{fig:phase_propa}.
The dashed lines show the (linear or circular) figures traced out by the electric
field vectors.
The curved arrows in the top row indicate the sense of rotation of
$\boldvec{E}_{\rm L}$ and $\boldvec{E}_{\rm R}$.
The curved arrows in the bottom row show the angles through which
$\boldvec{E}_{\rm L}$ and $\boldvec{E}_{\rm R}$ have rotated from their common
initial direction at the source, which is along the orientation of linear
polarization at the source.
These angles are exactly equal to the phases $\phi_{\rm L}$ and $\phi_{\rm R}$ of
the left and right modes, when the phase of a mode is defined through
Eq.~(\ref{eq_phase}).
Clearly, $\boldvec{E}_{\rm R}$ has rotated more than $\boldvec{E}_{\rm L}$, so the
orientation of linear polarization has rotated in the sense of the right mode.
Hence, Faraday rotation is right-handed about $\boldvec{B}$.
%%%KF%%% After the corrected Figure 8.1 of \cite{rybicki&l_79}.
After the corrected Figure 8.1 of Rybicki \& Lightman (1979).
}
\label{fig:faraday_rotation}
\end{figure*}

At a distance $s$ from the source, the superposition of the right and left modes
still forms a linearly polarized wave.
However, because the right and left modes now have different phases ($\phi_{\rm R}
\not= \phi_{\rm L}$),
their electric field vectors, $\boldvec{E}_{\rm R}$ and $\boldvec{E}_{\rm L}$,
have rotated through different angles;
as a result, the electric field vector of the linearly polarized wave,
$\boldvec{E}_\ell$, has rotated with respect to its initial direction at the
source, $\boldhat{e}_\star$
(see Figure~\ref{fig:faraday_rotation}).
The associated rotation of the polarization orientation is known as {\it Faraday
rotation}.
This rotation occurs in the sense that the phase of the leading mode increases,
i.e., the sense that $\boldvec{E}_{\rm R}$  rotates,
which is the right-handed sense about $\boldvec{B}$.
The angle through which the polarization orientation rotates
is half the phase difference between the right and left modes (see
Appendix~\ref{sec:Efield_expression} for the exact derivation):
\begin{eqnarray}
\Delta \psi^{\scriptscriptstyle [B]}
& = & {\textstyle \frac{1}{2}} \ \Delta \phi
\nonumber \\
& = & \frac{1}{2} \ \int_0^s
\frac{\omega_{\rm e}^2 \, |\Omega_{\rm e}|}{\omega^3} \ k \ ds'
\nonumber \\
& = & \left( \frac{e^3}{2\pi \, m_{\rm e}^2 \, c^4} \
\int_0^s n_{\rm e} \ B \ ds' \right) \ \lambda^2 \ ,
\label{eq_delta_psiB}
\end{eqnarray}
where we have used Eq.~(\ref{eq_phase_diff}) for the second equality
and $\omega_{\rm e}^2 = \frac{\displaystyle 4\pi n_{\rm e} e^2}{\displaystyle
m_{\rm e}}$,
$|\Omega_{\rm e}| = \frac{\displaystyle e B}{\displaystyle m_{\rm e} c}$,
$\omega = c \, k$, and $k = \frac{\displaystyle 2\pi}{\displaystyle \lambda}$ for
the third equality.
Superscript ${\scriptstyle [B]}$ indicates that $\psi^{\scriptscriptstyle [B]}$ is
an angle about the magnetic field, $\boldvec{B}$ (measured in a right-handed
sense)
-- as opposed to $\psi$, which is an angle in the plane of the sky (measured
counterclockwise; see Figure~\ref{fig:IAU}).

From the perspective of the observer, located at a distance $d$ from the source,
the Faraday rotation angle over the entire path from the source ($s=0$) to the
observer ($s=d$) is
\begin{eqnarray}
\Delta \psi
& = & \pm \, \Delta \psi^{\scriptscriptstyle [B]} (s=d)
\nonumber \\
& = & \left( \frac{e^3}{2\pi \, m_{\rm e}^2 \, c^4} \
\int_0^d n_{\rm e} \ (\pm B) \ ds \right) \ \lambda^2 \ ,
\label{eq_delta_psi}
\end{eqnarray}
with the $+$ [$-$] sign applying if $\boldvec{B}$ points toward [away from] the
observer.

Eq.~(\ref{eq_delta_psi}) was derived in the case of wave propagation parallel to
the magnetic field, $\boldvec{B}$.
It can be shown that this equation remains valid for other propagation directions
provided the factor $(\pm B)$ in the integral be replaced by the magnetic field
component in the (positive) propagation direction, $\boldvec{B} \! \cdot \!
\boldhat{e}_k$.
This field component is equivalent to the field component along the line of sight,
$B_\parallel$, as defined in the Faraday rotation community, which considers that
$B_\parallel$ is positive [negative] if $\boldvec{B}$ points toward [away from]
the observer.
Eq.~(\ref{eq_delta_psi}) can then be recast in the form
\begin{equation}
\Delta \psi \ = \ {\rm RM} \ \lambda^2 \ ,
\label{eq_rot_angle}
\end{equation}
where
\begin{equation}
{\rm RM} \ \equiv \
\frac{e^3}{2\pi \, m_{\rm e}^2 \, c^4} \
\int_0^d n_{\rm e} \ B_\parallel \ ds
\label{eq_RM}
\end{equation}
is the so-called rotation measure.
Note that the convention used here for the sign of $B_\parallel$ 
was precisely chosen to match the sign of ${\rm RM}$ \citep{manchester_72}.

Finally, the observed polarization angle of the incoming radiation, $\psi_{\rm
obs}$, is related to the intrinsic polarization angle at the source, $\psi_{\rm
src}$, and to the Faraday rotation angle, $\Delta \psi$, through
\begin{equation}
\psi_{\rm obs}
\ = \ \psi_{\rm src} + \Delta \psi
%%\ = \ \psi_{\rm src} + \ {\rm RM} \ \lambda^2
\ \cdot
\label{eq_pol_angle}
\end{equation}

To summarize, what we have learnt in this section
is that Faraday rotation is right-handed about the magnetic field, $\boldvec{B}$.
The implication for an observer looking toward a source of linearly polarized
synchrotron radiation is that Faraday rotation appears counterclockwise in the
plane of the sky if $\boldvec{B}$ points toward the observer and clockwise if
$\boldvec{B}$ points away from the observer.
Those are physical results, independent of any particular convention regarding,
e.g., the definition of the right and left circularly polarized modes
(Sect.~\ref{sec:confusion_modes}), the definition of the phase
(Sect.~\ref{sec:confusion_phase}), or the sign of $B_\parallel$
(Sect.~\ref{sec:confusion_Bparallel}).

\section{Possible sources of confusion}
\label{sec:confusion}

\subsection{Right and left modes}
\label{sec:confusion_modes}

Circularly polarized waves are divided into right and left modes,
based on the sense of rotation of the electric field vector, $\boldvec{E}$.
However, the reference vector about which the sense of rotation is measured can be
chosen in different ways (see summary in Table~\ref{table_rotation}).

For plasma physicists, the right (R) and left (L) circularly polarized modes
are defined with respect to the magnetic field, $\boldvec{B}$,
such that the electric field vector of the R [L] mode,
$\boldvec{E}_{\rm R}$ [$\boldvec{E}_{\rm L}$],
rotates in a right-handed [left-handed] sense about $\boldvec{B}$
\citep[see, e.g.,][]{nicholson_83, chen_16}.

For astronomers, the right circularly polarized (RCP)
and left circularly polarized (LCP) modes
are defined with respect to the wave vector, $\boldvec{k}$,
i.e., from the perspective of an observer looking at the sky
and watching the wave approach.
Radio astronomers use the IEEE convention,
according to which the electric field vector of the RCP [LCP] mode,
$\boldvec{E}_{\rm \scriptscriptstyle RCP}$ [$\boldvec{E}_{\rm \scriptscriptstyle
LCP}$],
rotates in a right-handed [left-handed] sense about $\boldvec{k}$,
i.e., counterclockwise [clockwise] as viewed by the observer.
Optical astronomers use the opposite convention
\citep[see, e.g.,][]{robishaw&h_18}.

\begin{figure*}
\begin{minipage}{\textwidth}
\centering
\includegraphics[width=\textwidth]{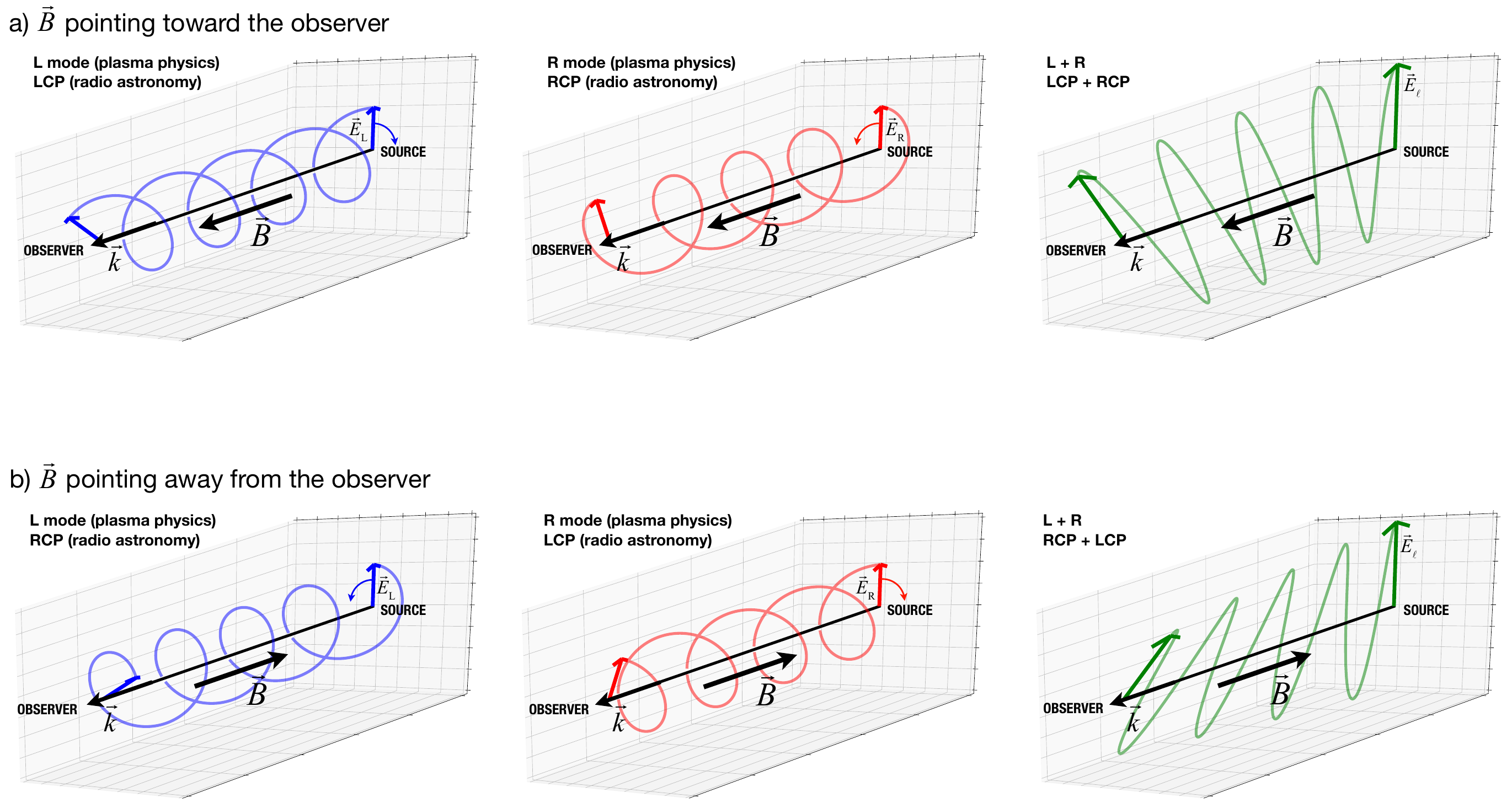}
\caption{
3D patterns formed by the tips of the electric field vectors
of a linearly polarized radio wave (right panels)
and of its left and right circularly polarized components,
defined either with respect to the magnetic field, $\boldvec{B}$,
as the L and R modes (plasma physics convention),
or with respect to the wave vector, $\boldvec{k}$,
as the LCP and RCP modes (radio astronomy convention).
The electric field vector of the linearly polarized wave, $\boldvec{E}_\ell$,
defines the orientation of linear polarization (right panels).
The latter undergoes Faraday rotation$^{\rm a}$ between the source (far side) and
the observer (near side),
and this rotation has the same handedness as the propagation helix of the L mode
(left panels).
(a) $\boldvec{B}$ points toward the observer,
so the L [R] mode corresponds to the LCP [RCP] mode
and Faraday rotation is right-handed about $\boldvec{k}$,
i.e., counterclockwise for the observer.
(b) $\boldvec{B}$ points away from the observer,
so the L [R] mode now corresponds to the RCP [LCP] mode
and Faraday rotation is left-handed about $\boldvec{k}$,
i.e., clockwise for the observer.
An animated version of this plot, which more explicitly demonstrates the
propagation in time, is available as online supplementary material on the journal
website. The video can also be found at \url{https://youtu.be/rB_giDghgr4}. }
\label{fig:faraday_rotation_3D}
\parbox{\textwidth}
{
\begin{flushleft}
\scriptsize{
$^{\rm a}\ $ Here, for illustrative purposes, the Faraday rotation of the
polarization orientation has been hugely exaggerated with respect to the rotation
of the electric field vectors $\boldvec{E}_{\rm L}$ and $\boldvec{E}_{\rm R}$.
}
\end{flushleft}
}
\end{minipage}
\end{figure*}

Thus, in radio astronomy, the R [L] wave is seen as an RCP [LCP] wave
if $\boldvec{B}$ points toward the observer
(see Figure~\ref{fig:faraday_rotation_3D}a)
and as an LCP [RCP] wave if $\boldvec{B}$ points away from the observer
(see Figure~\ref{fig:faraday_rotation_3D}b).
Let us now examine the properties of the RCP and LCP modes in more detail.

Regardless of the direction of $\boldvec{B}$ with respect to the observer,
the temporal behavior (with increasing $t$) 
of $\boldvec{E}_{\rm \scriptscriptstyle RCP}$ [$\boldvec{E}_{\rm
\scriptscriptstyle LCP}$] at the source
is a right-handed [left-handed] rotation about $\boldvec{k}$.
This temporal behavior at the source propagates toward the observer,
such that the state at a distance $s$ from the source is delayed
with respect to the state at the source.
Hence, moving from the source toward the observer at a given time $t$
is equivalent to going backward in time at the source.
As a result, the spatial behavior (with increasing $s$)
of $\boldvec{E}_{\rm \scriptscriptstyle RCP}$ [$\boldvec{E}_{\rm
\scriptscriptstyle LCP}$] at a given time $t$
is a left-handed [right-handed] rotation about $\boldvec{k}$.
Stated differently, the tip 
of $\boldvec{E}_{\rm \scriptscriptstyle RCP}$ [$\boldvec{E}_{\rm
\scriptscriptstyle LCP}$]
at a given time $t$ traces out a left-handed [right-handed] helix
along the propagation direction.\footnote{
This explains the convention adopted by optical astronomers.
While radio astronomers base their definition of right and left circularly
polarized modes on the sense of rotation (in time) of $\boldvec{E}_{\rm
\scriptscriptstyle RCP}$ and $\boldvec{E}_{\rm \scriptscriptstyle LCP}$ at a given
position, optical astronomers rely on the handedness of the propagation helix (in
space) of $\boldvec{E}_{\rm \scriptscriptstyle RCP}$ and $\boldvec{E}_{\rm
\scriptscriptstyle LCP}$ at a given time.
}
This is illustrated in the middle panel of Figure~\ref{fig:faraday_rotation_3D}a
and left panel of Figure~\ref{fig:faraday_rotation_3D}b
[left panel of Figure~\ref{fig:faraday_rotation_3D}a
and  middle panel of Figure~\ref{fig:faraday_rotation_3D}b].
Note, in preparation for Sect.~\ref{sec:helicity},
that the magnetic field vector of each mode
describes the same propagation helix as the electric field vector,
with the implication that the associated current helicity
(defined in Sect.~\ref{sec:helicity}) is positive [negative].

The above considerations offer an alternative way of obtaining the sense of
Faraday rotation.
As we saw in Sect.~\ref{sec:mathderivation_circular}, the L mode has a slightly
slower phase velocity, and hence a slightly shorter wavelength
$\left( \lambda = 2\pi \frac{\displaystyle V_\phi}{\displaystyle \omega} \right)$,
than the R mode at the same angular frequency $\omega$ (see
Eq.~\ref{eq_phase_vel}).
Accordingly, the propagation helix of the L mode
(left panels in Figure~\ref{fig:faraday_rotation_3D})
goes through slightly more cycles between the source and the observer
than the propagation helix of the R mode (middle panels).
The effect is weak and can only be noticed on close examination of the figure.
The resulting Faraday rotation (right panels) occurs in the sense
that the helix with more cycles, i.e., the helix of the L mode,
winds up while approaching the observer,
which is the right-handed sense about $\boldvec{k}$
-- counterclockwise for the observer --
if $\boldvec{B}$ points toward the observer
(Figure~\ref{fig:faraday_rotation_3D}a)
and the left-handed sense about $\boldvec{k}$
-- clockwise for the observer --
if $\boldvec{B}$ points away from the observer
(Figure~\ref{fig:faraday_rotation_3D}b).
Clearly, this conclusion is identical to that reached in
Sect.~\ref{sec:mathderivation_linear},
based on the  phase difference, $\Delta \phi = \phi_{\rm R} - \phi_{\rm L}$.

An obvious pitfall here is to mistake the sense of rotation of a mode with
increasing time $t$, at a given distance $s$ from the source,
for the sense of rotation %%along the propagation helix 
with increasing distance $s$, at a given time $t$.
The possible confusion is discussed in some detail by
\cite{kliger&lr_90,shurcliff_63,goldstein_11}.
It is also illustrated in the animated version of Figure 4 (at
https://youtu.be/xXPEJjlcZ2w), which shows the propagation in time much more
clearly than the static figure.
The above mistake leads to the erroneous conclusion that the RCP [LCP] mode
describes a right-handed [left-handed] helix along the propagation direction,
which in turn yields the wrong sense of Faraday rotation.

\begin{table}
\begin{tabular}{ccccc}
\hline\hline
\noalign{\medskip}
Concerned & Electric & Sense of rotation & Sense of rotation & Reference \\
scientific & field & with increasing $t$ & with increasing $s$ & vector \\
community & vectors & at constant $s$ & at constant $t$ & or plane \\
\noalign{\medskip}
\hline
\noalign{\medskip}
Plasma & $\boldvec{E}_{\rm R}$ & \quad right-handed & \quad left-handed & \quad
about $\boldvec{B}$ \\
physics & & & & \\
& $\boldvec{E}_{\rm L}$ & \quad left-handed & \quad right-handed & \quad about
$\boldvec{B}$ \\
\noalign{\medskip}
\hline
\noalign{\medskip}
Radio & $\boldvec{E}_{\rm \scriptscriptstyle RCP}$ & \quad right-handed & \quad
left-handed & \quad about $\boldvec{k}$ \\
astronomy & & \quad or CCW & \quad or CW & \quad in the sky \\
\noalign{\medskip}
& $\boldvec{E}_{\rm \scriptscriptstyle LCP}$ & \quad left-handed & \quad
right-handed & \quad about $\boldvec{k}$ \\
& & \quad or CW & \quad or CCW & \quad in the sky \\
\noalign{\medskip}
\hline
\noalign{\medskip}
Optical & $\boldvec{E}_{\rm \scriptscriptstyle RCP}$ & \quad left-handed & \quad
right-handed & \quad about $\boldvec{k}$ \\
astronomy & & \quad or  CW & \quad or CCW & \quad in the sky \\
\noalign{\medskip}
& $\boldvec{E}_{\rm \scriptscriptstyle LCP}$ & \quad right-handed & \quad
left-handed & \quad about $\boldvec{k}$ \\
& & \quad or CCW & \quad or CW & \quad in the sky \\
\noalign{\medskip}
\hline\hline
\end{tabular}
\caption{
Definitions of the right (R or RCP) and left (L or LCP) circularly polarized modes
in different scientific communities (first column),
based on the sense of rotation of the associated electric field vectors (second
column),
both with increasing time $t$ at constant distance $s$ from the source (third
column)
and with increasing $s$ at constant $t$ (fourth column).
The reference vector about which the sense of rotation is measured (fifth column)
is either the magnetic field, $\boldvec{B}$ (in plasma physics), or the wave
vector, $\boldvec{k}$ (in radio astronomy and in optical astronomy).
Right-handed [left-handed] rotation about $\boldvec{k}$ is equivalent to
counterclockwise [clockwise] rotation in the plane of the sky.
The sense of rotation about $\boldvec{k}$ in the fourth column corresponds to the
handedness of the propagation helix in Figure~\ref{fig:faraday_rotation_3D}.
}
\label{table_rotation}
\end{table}

\subsection{Phase of a wave}
\label{sec:confusion_phase}

The phase of a wave can also be defined through the opposite of
Eq.~(\ref{eq_phase}),
namely,
\begin{equation}
\phi \ = \ \int_0^s k \, ds' - \omega\,t \ \cdot
\label{eq_phase_bis}
\end{equation}
Let us see how proceeding from Eq.~(\ref{eq_phase_bis}) instead of
Eq.~(\ref{eq_phase})
affects the derived sense of Faraday rotation.
Consider again a linearly polarized radio wave with angular frequency $\omega$
propagating parallel to the magnetic field, $\boldvec{B}$.
As before, this wave can be decomposed into right (R) and left (L) circularly
polarized modes with the same $\omega$.

At the source ($s=0$), the right and left modes have again the same phase,
$\phi_\star(t)$, but now $\phi_\star(t) = - \omega\,t$.
At a distance $s$ from the source, the two modes have a phase difference,
$\Delta \phi \equiv \phi_{\rm R} - \phi_{\rm L}$, which is now given by
\begin{eqnarray}
\Delta \phi
& = & \int_0^s \Delta k \ ds'
\nonumber \\
& = & - \int_0^s \frac{\Delta V_\phi}{V_\phi} \ k \ ds'
\nonumber \\
& = & - \int_0^s \frac{\omega_{\rm e}^2 \, |\Omega_{\rm e}|}{\omega^3} \ k \ ds' \
,
\label{eq_phase_diff_bis}
\end{eqnarray}
instead of Eq.~(\ref{eq_phase_diff}).
This phase difference is now negative, which means that, at any distance $s$ from
the source,
the right mode is less advanced in phase than the left mode.

There are two ways to interpret this result physically,
both based on the right mode having the larger phase velocity
($V_{\phi,{\rm R}} > V_{\phi,{\rm L}}$).
The first interpretation is analogous to that proposed in
Sect.~\ref{sec:mathderivation_linear}, below Eq.~(\ref{eq_phase_diff}).
At the source, the right and left modes at any time $t_\star$ have the same phase,
$\phi_\star(t_\star) = -\omega\,t_\star$, which is now a decreasing function of
$t_\star$.
The right mode is faster to propagate this decreasing $\phi_\star(t_\star)$ out to
a given distance $s$ from the source,
so that the phase of the right mode at distance $s$ is smaller (or more negative)
than the phase of the left mode.

Alternatively, we can identify the phase of a mode, which is now an increasing
function of $s$,
with the angle through which the electric field vector turns 
along the propagation helix
(see middle [left] panels in Figure~\ref{fig:faraday_rotation_3D} for the right
[left] mode),
starting from the point corresponding to the initial direction at the source
and moving in the propagation direction (i.e., in a left-handed [right-handed]
sense about $\boldvec{B}$ for the right [left] mode).
Since the helix of the left mode winds up at a faster rate than the helix of the
right mode (see Sect.~\ref{sec:confusion_modes}),
the phase of the left mode at a given distance $s$ is larger than the phase of the
right mode.

In both cases, the left mode is ahead in phase relative to the right mode.
The first interpretation is more directly relevant for times greater than the
phase travel times to a given distance $s$ ($t > \frac{\displaystyle
s}{\displaystyle V_{\phi,{\rm L}}} > \frac{\displaystyle s}{\displaystyle
V_{\phi,{\rm R}}}$),
whereas the second interpretation is easier to picture for distances greater than
the phase travel distances on a given time $t$ ($s > V_{\phi,{\rm R}} \, t >
V_{\phi,{\rm L}} \, t$).
However, both interpretations can be generalized to all times and distances.

The consequence for the linearly polarized wave
is again that the polarization orientation rotates
in the sense that the phase of the leading mode increases,
which is now the sense that the propagation helix of the left mode
winds up while approaching the observer.
Regardless of whether $\boldvec{B}$ points toward the observer
(Figure~\ref{fig:faraday_rotation_3D}a)
or away from the observer (Figure~\ref{fig:faraday_rotation_3D}b),
this is again the right-handed sense about $\boldvec{B}$.
Thus, Faraday rotation remains right-handed about the magnetic field.

\subsection{Two examples of incorrect derivation}
\label{sec:confusion_examples}

To illustrate the subtleties and pitfalls 
discussed in Sects.~\ref{sec:confusion_modes} and \ref{sec:confusion_phase},
we now critically examine two derivations presented in widely read textbooks,
where a subtle error in the sense of rotation of the electric field vectors of the
right and left modes, $\boldvec{E}_{\rm R}$ and $\boldvec{E}_{\rm L}$, has led to
the wrong sense of Faraday rotation.

Let us first look at Chapter 4 of the plasma textbook by \cite{chen_16}.
There, the phase of a mode is defined through an equation (Eq.~4.7) similar to our
Eq.~(\ref{eq_phase_bis}), stating that the phase increases with increasing
distance from the source.
In his Sect.~4.17.2, Chen argues that the left mode undergoes more cycles over a
given distance than the right mode, from which he correctly concludes that, at a
distance $d$ from the source, the left mode is more advanced in phase than the
right mode.
Besides, he implicitly identifies the phase of the right [left] mode with the
angle through which $\boldvec{E}_{\rm R}$ [$\boldvec{E}_{\rm L}$] turns upon
traversing the plasma, i.e., in our terminology, the angle through which
$\boldvec{E}_{\rm R}$ [$\boldvec{E}_{\rm L}$] turns along the propagation helix
between $s=0$ and $s=d$.

The problem is that Chen takes $\boldvec{E}_{\rm R}$ [$\boldvec{E}_{\rm L}$] to
turn (in space) along the propagation helix %%at a given time 
in the same sense as $\boldvec{E}_{\rm R}$ [$\boldvec{E}_{\rm L}$] rotates (in
time) at a given position, i.e., in a right-handed [left-handed] sense about
$\boldvec{B}$ (see his Figure~4.44), while in reality $\boldvec{E}_{\rm R}$
[$\boldvec{E}_{\rm L}$] turns in the opposite sense
(see middle [left] panels in our Figure~\ref{fig:faraday_rotation_3D}).
This error leads him to incorrectly conclude that, at a distance $d$ from the
source, $\boldvec{E}_{\rm L}$ has turned more in a left-handed sense than
$\boldvec{E}_{\rm R}$ in a right-handed sense, and, by implication, that Faraday
rotation is left-handed about the magnetic field.

A similar mistake was probably made in Sect.~8.2 of the astrophysics textbook by
\cite{rybicki&l_79}.
Their exact line of thought is more difficult to follow, 
because the derivation is very compact, 
some key information is missing, and unfortunately a critical sign error is
present.
To start with, Rybicki \& Lightman adopt the optical convention for the right and
left modes, according to which the electric field vector of the right [left] mode,
$\boldvec{E}_{\rm R'}$ [$\boldvec{E}_{\rm L'}$],\footnote{
We added a prime to the indices ${\rm R}$ and ${\rm L}$ of \cite{rybicki&l_79} to
distinguish their definition of right and left modes from ours.
}
rotates in a left-handed [right-handed] sense about the wave vector,
$\boldvec{k}$.
They define the phase of a mode through the same equation as \cite{chen_16},
similar to our Eq.~(\ref{eq_phase_bis}).
Based on these two premises, the time variation of $\boldvec{E}_{\rm R'}$
[$\boldvec{E}_{\rm L'}$] is described by their Eq.~(8.24) with the upper [lower]
sign.\footnote{
In reality, the $\pm$ sign in Eq.~(8.24) should be replaced by a $\mp$ sign.
}
Since the ambient magnetic field, $\boldvec{B}_0$ (their Eq.~8.25), is assumed to
point in the same direction as $\boldvec{k}$ (toward the reader in their Figure
8.1), $\boldvec{E}_{\rm R'}$ [$\boldvec{E}_{\rm L'}$] corresponds to
$\boldvec{E}_{\rm L}$ [$\boldvec{E}_{\rm R}$] in our notation (see
Table~\ref{table_rotation}).

The dispersion relation (their Eq.~8.27) is given 
in the form of an equation for the dielectric constant, $\epsilon =
\frac{\displaystyle c^2 k^2}{\displaystyle \omega^2}$ (their Eq.~8.9),
with the plasma frequency, $\omega_p$ (their Eq.~8.11), equivalent to our
$\omega_{\rm e}$ and the cyclotron frequency, $\omega_B$ (their Eq.~8.21),
equivalent to our $|\Omega_{\rm e}|$.
It is easily verified that their Eq.~(8.27) is equivalent to our
Eq.~(\ref{eq_DR}), with the right and left modes swapped.
Their Eq.~(8.27) then leads to an expression for the wavenumber, $k$ (their
Eq.~8.29), which implies $k_{\rm R'} > k_{\rm L'}$, such that $\boldvec{E}_{\rm
R'}$ turns (in space) at a faster rate than $\boldvec{E}_{\rm L'}$.

For illustration, Figure 8.1 of \cite{rybicki&l_79} shows the electric field
vectors at two different positions (say, $s=0$ in Fig.~8.1a and $s=d$ in
Fig.~8.1b), with $\boldvec{E}_{\rm R'}$ (left panel) [$\boldvec{E}_{\rm L'}$
(middle panel)] rotating (in time) in a left-handed [right-handed] sense about
$\boldvec{B}_0$.
The sense that $\boldvec{E}_{\rm R'}$ [$\boldvec{E}_{\rm L'}$] turns from $s=0$ to
$s=d$ is not clearly indicated in the figure, but since $\boldvec{E}_{\rm R'}$
must turn through a larger angle than $\boldvec{E}_{\rm L'}$ (as implied by
$k_{\rm R'} > k_{\rm L'}$), we gather that $\boldvec{E}_{\rm R'}$
[$\boldvec{E}_{\rm L'}$] turns in a left-handed [right-handed] sense about
$\boldvec{B}_0$, consistent with Figure 8.1b (right panel) showing left-handed
Faraday rotation.

This again is incorrect, and again the error comes from mistaking the sense that
$\boldvec{E}_{\rm R'}$ [$\boldvec{E}_{\rm L'}$] rotates (in time)
at a given position for the sense that $\boldvec{E}_{\rm R'}$ [$\boldvec{E}_{\rm
L'}$] turns (in space) in the propagation direction at a given time.

\subsection{Sign of $B_\parallel$}
\label{sec:confusion_Bparallel}

Two opposite conventions for the sign of the magnetic field component along the
line of sight, $B_\parallel$, are being used by radio astronomers \citep[see,
e.g.,][]{robishaw&h_18}.

In the Zeeman splitting community, 
the sign convention for $B_\parallel$ is analogous to 
the sign convention adopted for the line-of-sight velocity,
$v_\parallel$, measured through the Doppler effect.
In the same way as $v_\parallel$ is taken to be positive for motions away from the
observer,
$B_\parallel$ is taken to be positive when the magnetic field, $\boldvec{B}$,
points away from the observer
\citep{verschuur_69}.

In the Faraday rotation community, the opposite sign convention is used, namely,
$B_\parallel$ is taken to be positive when $\boldvec{B}$ points toward the
observer.
This convention dates back to \cite{manchester_72}, who chose to have the sign of
$B_\parallel$ match the sign of ${\rm RM}$ (see
Sect.~\ref{sec:mathderivation_linear}).

The only impact of the chosen convention for the sign of $B_\parallel$
is in the expression of ${\rm RM}$.
With the sign convention from the Zeeman splitting community,
$B_\parallel$ would have the opposite sign to ${\rm RM}$,
which would then be given by
\begin{equation}
{\rm RM} \ = \
- \frac{e^3}{2\pi \, m_{\rm e}^2 \, c^4} \
\int_0^d n_{\rm e} \ B_\parallel \ ds \ \cdot
\nonumber
\end{equation}
But again, the derived sense of Faraday rotation would remain unchanged.

The existence of opposite conventions for the sign of $B_\parallel$ could
potentially be a source of error.
\cite{green&mcrh_12} compared the magnetic field directions inferred from Zeeman
splitting of 18-cm OH masers in Galactic star-forming regions with those inferred
from Faraday rotation of Galactic pulsars and extragalactic radio sources along
nearby lines of sight.
Both kinds of measurements indicate coherent magnetic field directions across the
observed region, but the field directions inferred from Zeeman splitting and from
Faraday rotation are in opposition.
Since the two methods probe different media, the discrepencay could very well have
a physical origin.
However, it could also be that an error was made either in the Zeeman splitting
convention (as discussed by the authors) or in the Faraday rotation convention.

\subsection{Limits of integration}
\label{sec:confusion_limits}

There has been some confusion in the literature regarding the order 
of the integration limits in the expression of the rotation measure.
Our derivation in Sect.~\ref{sec:mathderivation_linear} led us to write ${\rm RM}$
as an integral from the source ($s=0$) to the observer ($s=d$):
\begin{equation}
{\rm RM} \ = \ {\cal C} \ \int_{s=0}^{s=d} n_{\rm e} \ B_\parallel \ ds \ ,
\label{eq_RM_red}
\end{equation}
where ${\cal C} \equiv \frac{\displaystyle e^3}{\displaystyle 2\pi \, m_{\rm e}^2
\, c^4}$
is the prefactor appearing in the right-hand side of Eq.~(\ref{eq_RM}),
and $B_\parallel$ is the line-of-sight component of the magnetic field,
$\boldvec{B}$,
taken to be positive when $\boldvec{B}$ points from the source to the observer.
In view of this convention for the sign of $B_\parallel$,
Eq.~(\ref{eq_RM_red}) can be recast in vectorial form:
\begin{equation}
{\rm RM} \ = \ {\cal C} \ \int_{\rm src}^{\rm obs} n_{\rm e} \
\boldvec{B} \cdot d\boldvec{s} \ \cdot
\label{eq_RM_vec}
\end{equation}

\begin{figure}
\centering
\includegraphics[width=0.47\textwidth]{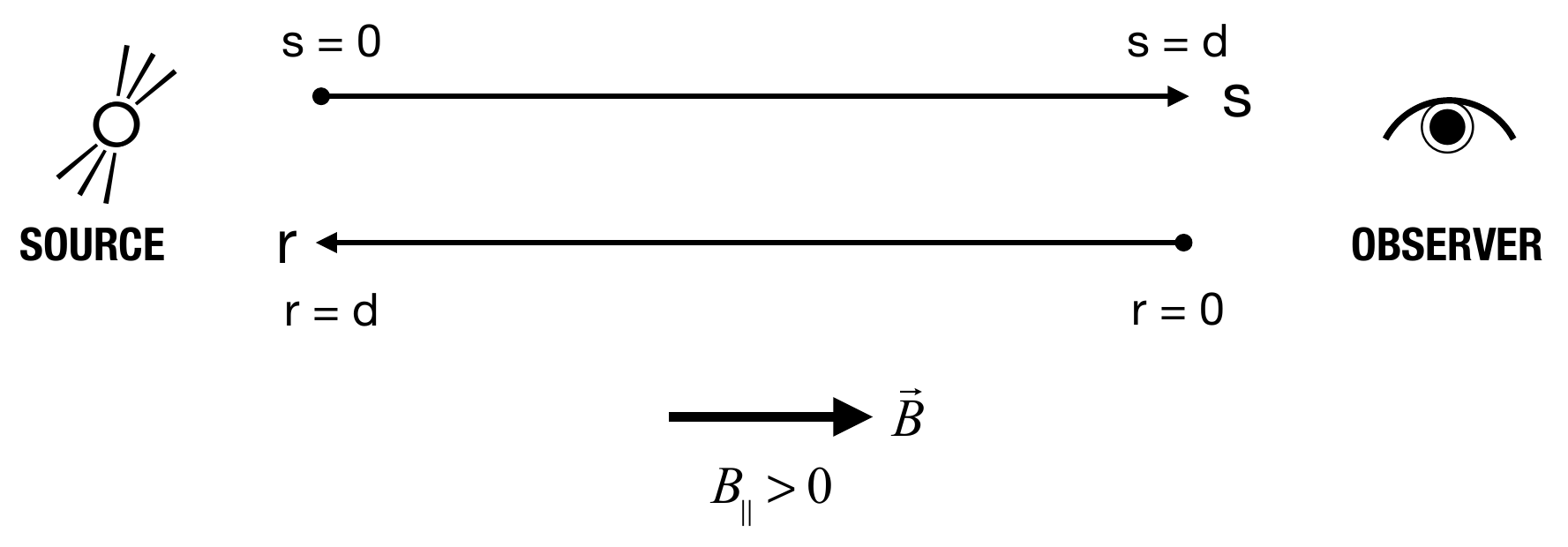}
\caption{
The two complementary coordinates used in this paper to locate an arbitrary point
between a radio source and an observer:
$s$ is the distance from the source along the propagation direction
and $r$ is the distance from the observer along the line of sight.
Also shown is the convention for the sign of $B_\parallel$ used in the Faraday
rotation community: $B_\parallel > 0$ when the magnetic field points from the
source to the observer, i.e., in the opposite direction to increasing $r$.
}
\label{fig:los_coordinates}
\end{figure}

In practice, it is often more convenient to place the origin
of the line-of-sight coordinate at the observer, rather than at the source.
If we denote by $r$ the line-of-sight distance from the observer
(see Figure~\ref{fig:los_coordinates}), such that $r = d - s$,
we can rewrite Eq.~(\ref{eq_RM_red}) as an integral over $r$.
Noting that $s=0$ corresponds to $r=d$ (at the source),
$s=d$ corresponds to $r=0$ (at the observer),
and $ds = -dr$, we obtain
\begin{equation}
{\rm RM} \ = \ {\cal C} \ \int_{r=d}^{r=0} n_{\rm e} \ B_\parallel \ (-dr) \ ,
\nonumber
\end{equation}
which is equivalent to
\begin{equation}
{\rm RM} \ = \ {\cal C} \ \int_{r=0}^{r=d} n_{\rm e} \ B_\parallel \ dr \ \cdot
\label{eq_RM_obs}
\end{equation}

Thus, the expression of ${\rm RM}$ can be written as an integral
either from the source ($s=0$) to the observer ($s=d$)
or from the observer ($r=0$) to the source ($r=d$),
with exactly the same integrand, $n_{\rm e} \, B_\parallel$.
The only requirement is that one sticks to the convention
that $B_\parallel$ is positive [negative] when $\boldvec{B}$
points toward [away from] the observer.
In particular, if $n_{\rm e}$ and $B_\parallel$ are both constant 
along the line of sight,
Eqs.~(\ref{eq_RM_red}) and (\ref{eq_RM_obs}) reduce to the same expression,
${\rm RM} = {\cal C} \, n_{\rm e} \, B_\parallel \, d$,
with the same sign.

Several authors have mistakenly rewritten Eq.~(\ref{eq_RM_vec}) in the form
\begin{equation}
{\rm RM} \ = \ {\cal C} \ \int_d^0 n_{\rm e} \
\boldvec{B} \cdot d\boldvec{s}
\hfill ({\rm a})
\nonumber
\end{equation}
\citep[e.g.,][]{vaneck&etal_17, vaneck_18} or
\begin{equation}
{\rm RM} \ = \ {\cal C} \ \int_{d}^{0} n_{\rm e} \ B_\parallel \ dr
\hfill ({\rm b})
\nonumber
\end{equation}
\citep[e.g.,][]{sun&lgc_15, dickey&ltw_19, thompson&ldm_19, stein&dbi_20}.
Clearly, the idea was to keep the order of the integration limits
from the source to the observer.
However, Eq.~({\rm a}) is ill-defined (the integration limits should be position
vectors, not scalars) and Eq.~({\rm b}) misses a minus sign, which arises from the
transformation
$\boldvec{B} \cdot d\boldvec{s} = B_\parallel \, ds = - B_\parallel \, dr$.
Without this minus sign, Eq.~({\rm b}) is equivalent to
\begin{equation}
{\rm RM} \ = \ - {\cal C} \ \int_{0}^{d} n_{\rm e} \ B_\parallel \ dr \ ,
\nonumber
\end{equation}
which, together with ${\cal C} > 0$, $n_{\rm e} \ge 0$, and $dr \ge 0$
($r$ increasing from $0$ to $d$), implies that ${\rm RM}$
has the opposite sign to (the line-of-sight averaged) $B_\parallel$
-- in contradiction with our adopted convention.
Note, however, that the mistake made here is purely formal, with no impact on the
derived sign of $B_\parallel$ or on the inferred magnetic field direction.

\section{Faraday rotation, synchrotron emission, and helicity}
\label{sec:helicity}

In Sects.~\ref{sec:mathderivation} and \ref{sec:confusion},
we considered the Faraday rotation of a background source
of linearly polarized synchrotron radiation
by a foreground magnetized plasma.
Here, we drop the notion of a background source
and assume instead that the magnetized plasma itself,
in addition to causing Faraday rotation,
is also the site of synchrotron emission.
Such diffuse synchrotron emission pervades the ISM
and is commonly used to study magnetic fields in the Milky Way Galaxy
\cite[e.g.,][]{haverkorn_15}.
Thus, we switch from a situation where synchrotron emission
and Faraday rotation are well separated in space
to a situation where they are spatially mixed.

Synchrotron emission is linearly polarized perpendicularly
to the local magnetic field projected onto the plane of the sky,
$\boldvec{B}_\perp$.
If we denote by $\psi_{B_\perp}$ the angle of $\boldvec{B}_\perp$ 
(measured counterclockwise from North; see Figure~\ref{fig:IAU}),
the intrinsic polarization angle of synchrotron emission is
\begin{equation}
\psi_{\rm src} \ = \ \psi_{B_\perp} \pm \frac{\pi}{2} \ \cdot
\label{eq_pol_angle_intr}
\end{equation}
As explained in Sect.~\ref{sec:mathderivation_linear},
the polarization angle of the emission produced at distance $r$ from the observer
undergoes Faraday rotation, changing from $\psi_{\rm src}(r)$ at the source
to $\psi_{\rm obs}(r)$ at the observer,
where, according to  Eqs.~(\ref{eq_pol_angle}), (\ref{eq_rot_angle}), and
(\ref{eq_RM_obs}),
\begin{eqnarray}
\psi_{\rm obs}(r)
& = & \psi_{\rm src}(r) + \Delta \psi(r)
\nonumber \\
& = & \psi_{\rm src}(r) + \Phi(r) \ \lambda^2
\label{eq_pol_angle_obs}
\end{eqnarray}
and
\begin{equation}
\Phi(r) \ \equiv \ {\cal C} \ \int_0^r n_{\rm e} \ B_\parallel \ dr' \ \cdot
\label{eq_FD}
\end{equation}
$\Phi(r)$ is the so-called Faraday depth\footnote{
Faraday depth is generally denoted by $\phi$ \citep{burn_66, gardner&w_66,
brentjens&d_05, vaneck_18}.
Here, we denote it by $\Phi$ to distinguish it from the phase of a mode defined
earlier (Eqs.~\ref{eq_phase} and \ref{eq_phase_bis}).
} 
at distance $r$ from the observer.
It has basically the same formal expression as ${\rm RM}$ in
Eq.~(\ref{eq_RM_obs}),
but is conceptually different:
whereas ${\rm RM}$ is a purely observational quantity,
which can be meaningfully defined only for a background synchrotron source,
$\Phi(r)$ is a truly physical quantity,
which can be defined at any point of the ISM,
independent of any background source.
The notion of Faraday depth is particularly useful in the present context,
where synchrotron emission and Faraday rotation are spatially mixed.

If the relativistic electrons responsible for synchrotron emission
have a power-law energy spectrum described by $f(E) = K_e \, E^{-\gamma}$,
the synchrotron emissivity at frequency $\nu$ is given by
\begin{equation}
{\cal E} \ = \
{\rm fc}(\gamma) \ K_e \ B_\perp^\frac{\gamma + 1}{2} \
\nu^{- \frac{\gamma - 1}{2}} \ ,
\label{eq_synchr_emiss}
\end{equation}
where ${\rm fc}(\gamma)$ is a known function of the electron spectral index,
and the intrinsic degree of linear polarization is
$p_{\rm src} = \frac{\gamma + 1}{\gamma + \frac{7}{3}}$
\citep[e.g.,][]{ginzburg&s_65}.
Then, for any given line of sight, the synchrotron {\it total} intensity is
\begin{equation}
I \ = \ \int_0^\infty {\cal E} \ dr
\label{eq_synchr_intensity}
\end{equation}
and the (complex) synchrotron {\it polarized} intensity is
\begin{equation}
P \ \equiv \ Q + i \, U
\ = \ \int_0^\infty p_{\rm src} \ {\cal E} \ e^{2 i \psi_{\rm obs}} \ dr \ \cdot
\label{eq_synchr_PI}
\end{equation}
The observed polarization angle is then given by
\begin{equation}
\psi = \frac{1}{2} \ \arctan \frac{U}{Q} \ ,
\label{eq_synchr_polangle}
\end{equation}
with $\arctan$ the two-argument arctangent function defined from $-180^\circ$ to
$+180^\circ$,
and the observed degree of linear polarization is given by
\begin{equation}
p = \frac{|P|}{I} = \frac{\sqrt{Q^2 + U^2}}{I} \ \cdot
\label{eq_synchr_polfrac}
\end{equation}

Because of the factor $e^{2 i \psi_{\rm obs}}$ in Eq.~(\ref{eq_synchr_PI}),
depolarization occurs when the synchrotron emission
produced at different distances along the line of sight
reaches the observer with different polarization angles.
According to Eq.~(\ref{eq_pol_angle_obs}), this happens 
either when the intrinsic polarization angle (Eq.~\ref{eq_pol_angle_intr})
varies along the line of sight
or when the Faraday depth (Eq.~\ref{eq_FD}) varies along the line of sight,
i.e., when Faraday rotation is present.
Eq.~(\ref{eq_pol_angle_obs}) also shows that Faraday rotation
by a magnetic field pointing toward the observer ($B_\parallel > 0$,
such that $\Phi(r)$ increases positively with increasing $r$)
acts in the same sense as counterclockwise rotation
of $\boldvec{B}_\perp$ away from the observer
(such that $\psi_{\rm src}(r)$ increases positively with increasing $r$).

\begin{figure*}
\centering
\includegraphics[width=\textwidth]{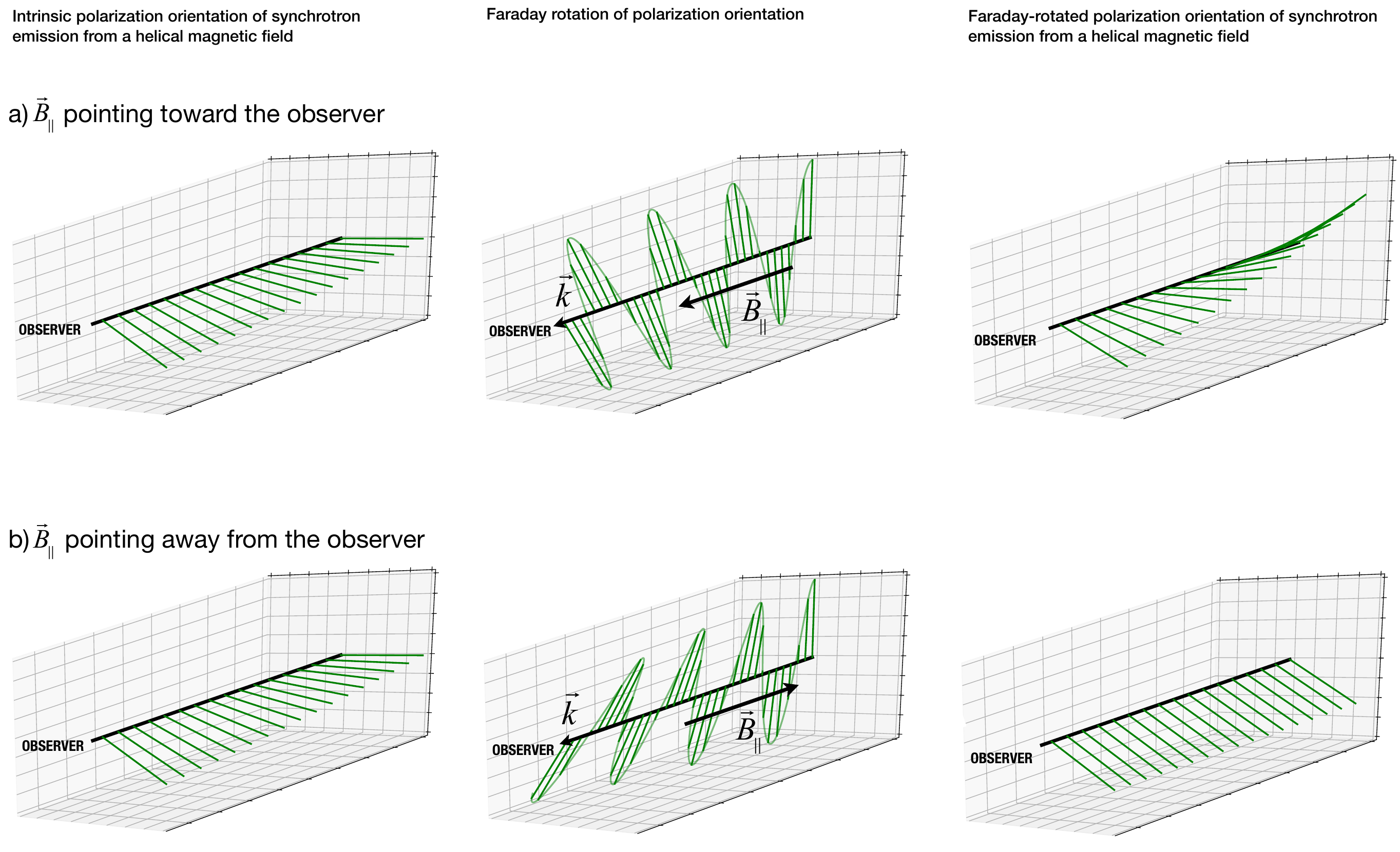}
\caption{
Cartoon illustrating how Faraday rotation can either reinforce or counteract
the depolarization effect of a helical magnetic field.
The perspective is the same as in Figure~\ref{fig:faraday_rotation_3D},
with the observer on the near side.
However, here, the background source is replaced by extended synchrotron emission 
throughout the Faraday-rotating plasma.
The magnetic field, $\boldvec{B}$, is described by Eq.~(\ref{eq_helical_mf}),
with $\frac{\displaystyle \partial \psi_{B_\perp}}{\displaystyle \partial Z} < 0$,
so the current helicity, given by Eq.~(\ref{eq_cur_helicity}), is positive ($H_j >
0$).
The \textbf{left panels} show the intrinsic polarization orientation of the
synchrotron emission 
produced at various distances $r$ from the observer
(first term in the right-hand side of Eq.~\ref{eq_pol_angle_obs}),
which is simply the orientation of the local $\boldvec{B}_\perp$ rotated by
$90^\circ$
(see Eq.~\ref{eq_pol_angle_intr}).
The \textbf{middle panels}, directly taken from the far-right column
in Figure~\ref{fig:faraday_rotation_3D},
show how the polarization orientation is Faraday-rotated
as the signal propagates along the line of sight
(second term in the right-hand side of Eq.~\ref{eq_pol_angle_obs}),
(a) when $\boldvec{B}_\parallel$ points toward the observer ($B_\parallel > 0$),
and (b) when $\boldvec{B}_\parallel$ points away from the observer ($B_\parallel <
0$).
The \textbf{right panels} show the polarization orientation of the synchrotron
emission 
produced at various distances $r$,
after Faraday rotation, i.e., as it reaches the observer
(sum of both terms in the right-hand side of Eq.~\ref{eq_pol_angle_obs}).
In scenario (a), the twisted polarization orientation from the helical field is
further wound by the Faraday rotation,
with the greatest amount of rotation at the back of the slab;
this yields a less coherent polarization signal,
equivalent to the signal from a more strongly helical field.
In scenario (b), the twisted polarization orientation from the helical field is
unwound by the Faraday rotation,
with again the greatest amount of rotation at the back of the slab;
this yields a strongly coherent polarization signal,
equivalent to the signal from a non-helical field.
}
\label{fig:helical_field}
\end{figure*}

A magnetic field with $\boldvec{B}_\perp$ rotating counterclockwise
away from the observer forms a "spiral-slide" helical magnetic field,
for which the tip of $\boldvec{B}_\perp$ traces out a left-handed helix 
(see Figure~8 in \cite{brandenburg&s_14} and top panels in
Figure~\ref{fig:helical_field}).
Such a helical magnetic field can be written in IAU coordinates
(see Figure~\ref{fig:IAU}) as
\begin{equation}
\boldvec{B} =
B_\perp \left( \cos \psi_{B_\perp} \ \boldhat{e}_X
+ \sin \psi_{B_\perp} \ \boldhat{e}_Y \right)
+ B_\parallel \ \boldhat{e}_Z \ ,
\label{eq_helical_mf}
\end{equation}
with $\frac{\displaystyle \partial \psi_{B_\perp}}{\displaystyle \partial Z} =
-\frac{\displaystyle \partial \psi_{B_\perp}}{\displaystyle \partial r} < 0$.
If $\boldvec{B}$ varies {\it only} along the line of sight,
the associated current helicity,
\begin{equation}
H_j
\ = \ \boldvec{B} \cdot \nabla \times \boldvec{B}
\ = \ - B_\perp^2 \ \frac{\partial \psi_{B_\perp}}{\partial Z} \ ,
\label{eq_cur_helicity}
\end{equation}
is positive, independent of the sign of $B_\parallel$.
Noting that $\frac{\displaystyle \partial \psi_{B_\perp}}{\displaystyle \partial
Z}
= - \frac{\displaystyle \partial \psi_{\rm src}}{\displaystyle \partial r}$,
Eq.~(\ref{eq_cur_helicity}) can be integrated to give
\begin{equation}
\psi_{\rm src}(r) - \psi_{\rm src}(0)
\ = \ \int_0^r \frac{H_j}{B_\perp^2} \ dr' \ \cdot
\label{eq_cur_helicity_int}
\end{equation}

The similarity between Eqs.~(\ref{eq_cur_helicity_int}) and (\ref{eq_FD})
reflects the similarity between the two depolarization mechanisms mentioned above.
On the one hand, the current helicity, $H_j$, of a simple spiral-slide helical
magnetic field
causes a line-of-sight variation of the intrinsic polarization angle, $\psi_{\rm
src}(r)$.
On the other hand, the line-of-sight magnetic field, $B_\parallel$,
in a magnetized plasma causes a line-of-sight variation of the Faraday depth,
$\Phi(r)$,
which indicates that Faraday rotation is occurring.

The two depolarization mechanisms can either reinforce or counteract each other,
as illustrated in Figures~\ref{fig:helical_field}a and \ref{fig:helical_field}b,
respectively.
If $H_j$ and $B_\parallel$ have the same sign,
Eqs.~(\ref{eq_cur_helicity_int}) and (\ref{eq_FD}) together
imply that $\psi_{\rm src}(r) - \psi_{\rm src}(0)$ and $\Phi(r)$
also have the same sign;
in that case, Faraday rotation amplifies the line-of-sight variation
of the intrinsic polarization angle, thereby increasing depolarization.
In contrast, if $H_j$ and $B_\parallel$ have opposite signs,
$\psi_{\rm src}(r) - \psi_{\rm src}(0)$ and $\Phi(r)$
also have opposite signs,
in which case Faraday rotation reduces the line-of-sight variation
of the intrinsic polarization angle, thereby decreasing depolarization.
In consequence, if the ambient magnetic field has positive [negative] current
helicity,
synchrotron emission from regions at positive [negative] Faraday depths
is strongly depolarized,
while synchrotron emission from regions at negative [positive] Faraday depths
is more weakly depolarized.

Nevertheless, there is an important difference between
the effect of the current helicity, $H_j$,
and the effect of the line-of-sight magnetic field, $B_\parallel$:
$H_j$ is intrinsic to the considered helical magnetic field,
and so is the handedness of the associated helix,
whereas $B_\parallel$ depends on the position of the observer,
and so does the sense -- or the effective handedness -- of the associated Faraday
rotation.

To illustrate this difference, let us revisit, and hopefully correct, 
the thought experiment described in the Appendix of \cite{brandenburg&s_14}.
Consider two hypothetical observers O$_1$ and O$_2$
located on opposite sides of a uniform plasma
in which both synchrotron emission and Faraday rotation take place.
Assume that the plasma is embedded in a simple spiral-slide helical magnetic
field,
with a line-of sight component pointing away from observer O$_2$ and toward
observer O$_1$.
Both observers see a helical magnetic field with the same $H_j$,
and hence the same handedness,
but they see opposite $B_\parallel$, and hence opposite Faraday rotation.
For observer O$_1$, $B_\parallel > 0$, Faraday rotation is counterclockwise
(see right panel in Figure~\ref{fig:faraday_rotation_3D}a),
the synchrotron-emitting region lies at positive Faraday depths ($\Phi>0$),
and the measured synchrotron emission is strongly [weakly] depolarized
if $H_j > 0$ [$H_j < 0$].
For observer O$_2$, $B_\parallel < 0$, Faraday rotation is clockwise
(see right panel in Figure~\ref{fig:faraday_rotation_3D}b),
the synchrotron-emitting region lies at negative Faraday depths ($\Phi<0$),
and the measured synchrotron emission is weakly [strongly] depolarized
if $H_j > 0$ [$H_j < 0$].

The idea that a helical magnetic field can compensate the depolarization effect
of Faraday rotation was first proposed by \cite{sokoloff&bsbbp_98}
and later studied through numerical simulations by \cite{volegova&s_10} and
\cite{brandenburg&s_14}.\footnote{
Both \cite{volegova&s_10} and \cite{brandenburg&s_14} consider magnetic helicity,
as opposed to current helicity,
but we believe that the conclusions they obtain with magnetic helicity
would remain qualitatively the same with current helicity.
Furthermore, \cite{volegova&s_10} and (in some places) \cite{brandenburg&s_14} use
the expression "rotation measure"
to refer to what we call "Faraday depth".
}
\cite{volegova&s_10} find that positive [negative] magnetic helicity leads to a
positive [negative]
correlation between rotation measure and polarization degree.
Similarly, \cite{brandenburg&s_14} conclude that
"positive [negative] magnetic helicity could be detected by observing
positive [negative] RM in highly polarized regions %%in the sky
and negative [positive] RM in weakly polarized regions.
Both conclusions are opposite to our own conclusion \citep{west&hfw_20}.
The reason is that the authors implicitly take Faraday rotation
to be left-handed about the magnetic field.
Although not directly stated in their papers, 
this can be inferred from two premises of their models:
(1) The $Z$-axis points from the observer to the source
(opposite to the IAU convention; see Figure~\ref{fig:IAU}).
Accordingly the polarization angle, $\psi$, and the Faraday rotation angle, $\Phi
\, \lambda^2$, increase clockwise for the observer.
In other words, the Faraday depth, $\Phi$, is positive
when Faraday rotation is clockwise for the observer.
(2) The Faraday depth is positive when the mean magnetic field
points toward the observer (standard convention).
Together, these two premises incorrectly imply that Faraday rotation
is left-handed about the magnetic field.\footnote{
\cite{brandenburg&s_14} take $B_\parallel = B_Z$, which implies that $B_\parallel$
is positive when the magnetic field points away from the observer (opposite to the
standard convention).
This is consistent with their Eq.~(3), which indicates that $\Phi$
has the opposite sign to $B_\parallel$.
}

\section{Conclusions}
\label{sec:conclusions}

Our paper aims to present a clear and unambiguous description of the true,
physical, sense of Faraday rotation, which we found lacking in existing texts.
To that end, we rederived the equations describing the propagation
of a linearly polarized radio electromagnetic wave undergoing Faraday rotation
in a magnetized plasma with magnetic field $\boldvec{B}$.
By decomposing the electric field vector of the linearly polarized wave,
$\boldvec{E}_\ell$, into the contributions from a right circularly polarized mode
(whose electric field vector, $\boldvec{E}_{\rm R}$, rotates in a right-handed
sense about $\boldvec{B}$) and a left circularly polarized mode (whose electric
field vector, $\boldvec{E}_{\rm L}$, rotates in a left-handed sense about
$\boldvec{B}$) and noting that the right mode has a slightly larger phase velocity
than the left mode,
we found that Faraday rotation is right-handed about $\boldvec{B}$.

This sense of Faraday rotation is easily understood physically.
In brief, Faraday rotation results from the action of $\boldvec{B}$
on the motion of free electrons accelerated by $\boldvec{E}_\ell$.
This action of $\boldvec{B}$, which occurs through the Lorentz force (second term
in the right-hand side of the momentum equation, Eq.~(\ref{eq_emomentum})),
is always in the sense of a right-handed rotation about $\boldvec{B}$
-- exactly as for the simple electron gyro-motion about magnetic field lines.
Since the motion of electrons produces an electric current,
which in turn acts back on the electric field (second term in the right-hand side
of Maxwell-Amp\`{e}re's equation, Eq.~(\ref{eq_maxwell_ampere})),
the right-handed rotation that $\boldvec{B}$ tends to impart to electrons
is automatically passed on to $\boldvec{E}_\ell$.

Our derived sense of Faraday rotation is by no means a new result.
It can be found in several textbooks \citep[e.g.,][]{stone_63, papas_65,
collett&s_12}.
But the opposite result -- that Faraday rotation is left-handed about
$\boldvec{B}$ --
is also found in the literature \citep[e.g.,][]{rybicki&l_79, chen_16}.
This has been the source of much confusion and uncertainty in the astrophysics
community.
Our recent work on the link between Faraday rotation and magnetic helicity
\citep{west&hfw_20} prompted us 
to produce a clear, systematic, and complete derivation,
which is both rigorous from the point of view of plasma physicists
and physically transparent for radio astronomers.
In that sense, our paper will help bridge any convention-related gap between
plasma physics and radio astronomy.

We also discussed the possible pitfalls that came to light
in the course of our investigation.
We often realized that an apparently correct and convincing reasoning
could lead to the wrong conclusion for a very subtle reason.
This is why we took great pains to explain all the steps of our derivation in
detail
and we carefully examined alternative approaches.

The most frequent errors and misconceptions that we encountered arise from a
confusion between the sense that the electric field vector of a mode rotates
with increasing time at a given distance from the source and the sense that the
electric field vector of this mode turns with increasing distance at a given time.
As a reminder (see Table~\ref{table_rotation}), at a given distance,
$\boldvec{E}_{\rm R}$ [$\boldvec{E}_{\rm L}$] rotates with increasing time in a
right-handed [left-handed] sense about $\boldvec{B}$, whereas at a given time,
$\boldvec{E}_{\rm R}$ [$\boldvec{E}_{\rm L}$] turns with increasing distance,
i.e., in the propagation direction, in a left-handed [right-handed] sense about
$\boldvec{B}$.
As explained in Sect.~\ref{sec:confusion_modes}, the reason why the two senses are
opposite is because moving away from the source at a given time is equivalent to
going backward in time at a given distance.

Similarly for radio astronomers, who define the right circularly polarized (RCP)
and left circularly polarized (LCP) modes with respect to the wave vector,
$\boldvec{k}$:
At a given position, the electric field vector of the RCP [LCP] mode,
$\boldvec{E}_{\rm \scriptscriptstyle RCP}$ [$\boldvec{E}_{\rm \scriptscriptstyle
LCP}$],
rotates in a right-handed [left-handed] sense about $\boldvec{k}$,
whereas at a given time, the propagation helix of the RCP [LCP] mode (plotted in
the middle panel of Figure~\ref{fig:faraday_rotation_3D}a and left panel of
Figure~\ref{fig:faraday_rotation_3D}b [left panel of
Figure~\ref{fig:faraday_rotation_3D}a and  middle panel of
Figure~\ref{fig:faraday_rotation_3D}b]) is left-handed [right-handed].

\begin{figure*}
\begin{minipage}{\textwidth}
\centering
\includegraphics{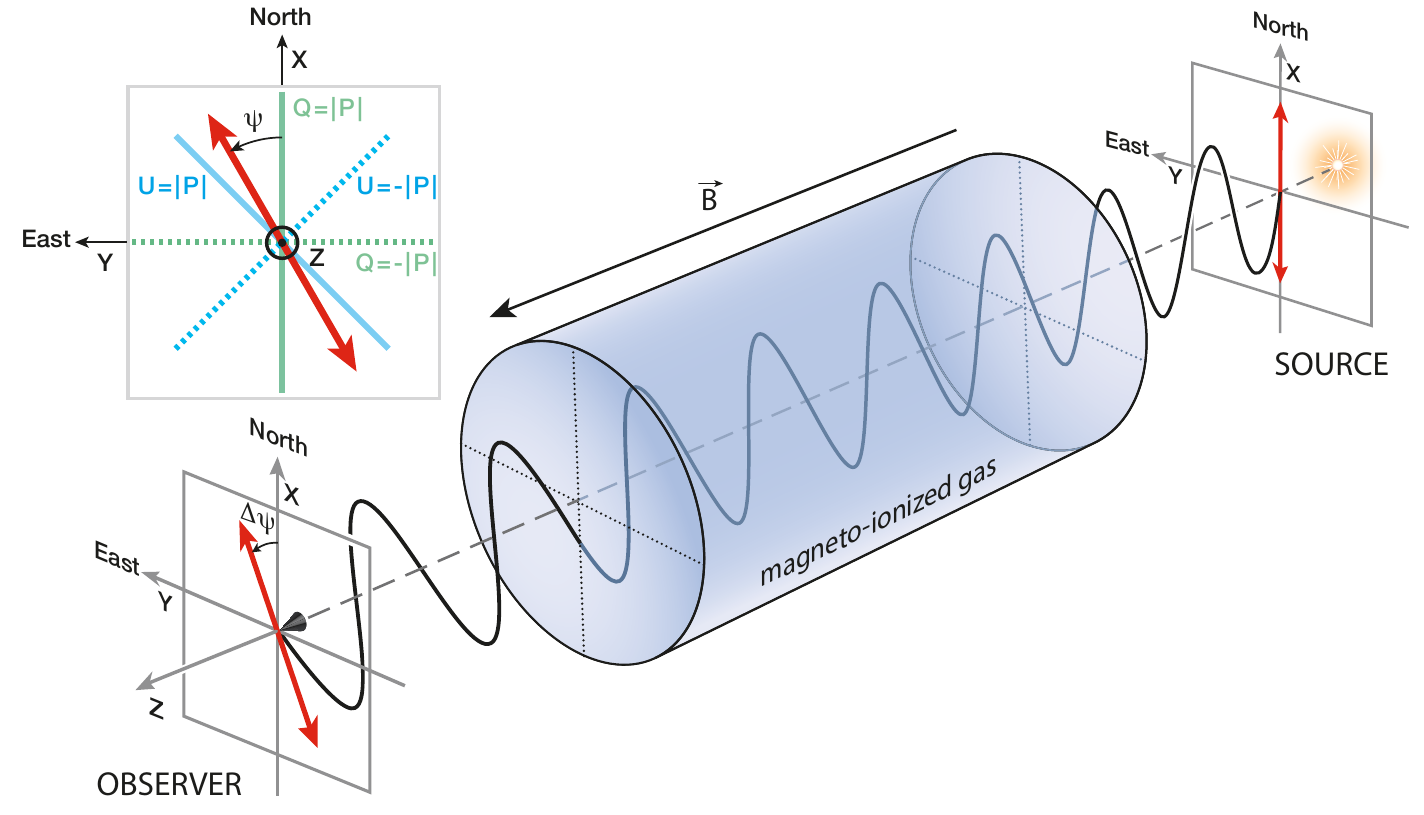}
\caption{
Summary plot showing qualitatively how the electric field vector of a linearly
polarized radio wave
oscillates along the propagation direction,
between the source (far side) and the observer (near side),
and how the polarization orientation (red double-headed arrow) undergoes Faraday
rotation as the wave passes through a magneto-ionized region (blue shaded
region).$^{\rm a}$
Faraday rotation is always right-handed about the magnetic field, $\boldvec{B}$.
When $\boldvec{B}$ points toward [away from] the observer,
Faraday rotation is counterclockwise [clockwise] in the plane of the sky;
this corresponds to a positive [negative] rotation angle, $\Delta \psi$
(Eq.~\ref{eq_delta_psi}), in the IAU definition
of the polarization angle (measured counterclockwise from North; see inset in the
upper-left corner).
For completeness, the inset also shows the axes of the Stokes parameters $Q$ and
$U$ (defined through Eq.~\ref{eq_synchr_PI}), i.e., the (faded green) lines where
$Q = \pm |P|$ (or $U = 0$) and the (faded blue) lines where $U = \pm |P|$ (or $Q =
0$).
Image credit:  NASA Goddard/Theophilus Britt Griswold.}
\label{fig:faraday_global}
\parbox{\textwidth}
{
\begin{flushleft}
\scriptsize{
$^{\rm a}\ $ Quantitatively, this plot is not realistic at all for the ISM, where
the Faraday rotation rate is tremendously smaller than the oscillation rate of the
electric field vector.
Typically, Faraday rotation occurs over parsec-scale distances, while the
wavelength of a radio wave is $\lambda = (30~{\rm cm}) \ \Big(\!
\frac{\displaystyle \nu}{\displaystyle 1~{\rm GHz}} \!\Big)^{-1}$.
}
\end{flushleft}
}
\end{minipage}
\end{figure*}

To conclude our paper, we present in Figure~\ref{fig:faraday_global} a summary
plot which clearly illustrates the physical process of Faraday rotation and
unambiguously shows the sense of rotation both with respect to the magnetic field
as well as from the perspective of the observer.
Since radio astronomers commonly use the Stokes parameters $Q$ and $U$ (defined
through Eq.~\ref{eq_synchr_PI}) to measure the sky, 
we include an inset in the upper-left corner, which shows the IAU coordinate and
polarization conventions, including the $Q$ and $U$ lines.
In this manner, Figure~\ref{fig:faraday_global} provides a complete "takeaway"
figure, which connects our theoretical discussion to observable quantities.

\section*{Acknowledgements}
Our paper was triggered by discussions about helicity (in particular, with Anvar
Shukurov) at the IMAGINE workshop of April, 2019.
We express our gratitude to the colleagues we contacted
in the course of our investigation
(Axel Brandenburg, Jo-Anne Brown, Gabriel Fruit, JinLin Han, George Heald,
Anna Ordog, Rodion Stepanov)
for their replies to our initial query
{\it "What is the correct sense of Faraday rotation?"},
for their willingness to engage in a longer discussion with us,
and for our constructive exchanges.
We also thank Theophilus Britt Griswold for his great help in producing
Figures~\ref{fig:phase_propa} and \ref{fig:faraday_global} and Bryan Gaensler for
his contributions to improving Figure~\ref{fig:faraday_global}.
J.L.W. acknowledges support of the Dunlap Institute, which is funded through an
endowment established by the David Dunlap family and the University of Toronto.
This research has made use of the NASA Astrophysics Data System (ADS).
Finally, we extend our sincere thanks to the referee for their extremely
constructive and detailed report, which helped us considerably improve the clarity
and readability of our paper.
%%%%%%%%%%%%%%%%%%%%%%%%%%%%%%%%%%%%%%%%%%%%%%%%%%

\section*{Data Availability}
No new data were generated or analysed in support of this research.



\bibliographystyle{mnras}
\bibliography{references} % if your bibtex file is called example.bib

%\begin{thebibliography}{99}
%\end{thebibliography}

%%%%%%%%%%%%%%%%%%%%%%%%%%%%%%%%%%%%%%%%%%%%%%%%%%

%%%%%%%%%%%%%%%%% APPENDICES %%%%%%%%%%%%%%%%%%%%%

\appendix

\section{Dispersion relation of parallel electromagnetic waves}
\label{sec:dispersion_relation}

The equations governing the propagation of an electromagnetic wave in a cold,
magnetized plasma with electron density $n_{\rm e}$
and magnetic field $\boldvec{B}_0 = B_0 \, \boldhat{e}_z$\footnote{
In the appendices, the ambient magnetic field is denoted by $\boldvec{B}_0$, to
distinguish it from the wave magnetic field vector, which is called $\boldvec{B}$
by analogy with the wave electric field vector, $\boldvec{E}$.
In the rest of the paper, where the wave magnetic field vector is not discussed,
the ambient magnetic field is denoted by $\boldvec{B}$.
There is also a distinction between the unit vector $\boldhat{e}_z$ in the
direction of $\boldvec{B}_0$ (used in the appendices) and the unit vector
$\boldhat{e}_Z$ in the direction of the wave vector, $\boldvec{k}$ (shown in
Figure~\ref{fig:IAU} and used in the rest of the paper).
} 
are the linearized evolution equations for the wave electric field vector,
$\boldvec{E}$, the wave magnetic field vector, $\boldvec{B}$, and the electron
velocity, $\boldvec{V}_{\rm e}$, i.e., Maxwell-Amp\`{e}re's equation,
\begin{equation}
\frac{\partial \boldvec{E}}{\partial t}
\ = \ c \ \nabla \times \boldvec{B} - 4\pi \ \boldvec{j} \ ,
\label{eq_maxwell_ampere}
\end{equation}
with the electric current density $\boldvec{j} = q_{\rm e} \, n_{\rm e} \,
\boldvec{V}_{\rm e}$,
Maxwell-Faraday's equation,
\begin{equation}
\frac{\partial \boldvec{B}}{\partial t}
\ = \ - c \ \nabla \times \boldvec{E} \ ,
\label{eq_maxwell_faraday}
\end{equation}
and the electron momentum equation,
\begin{equation}
m_{\rm e} \ \frac{\partial \boldvec{V}_{\rm e}}{\partial t}
\ = \ q_{\rm e} \ \left( 
\boldvec{E} + \frac{1}{c} \ \boldvec{V}_{\rm e} \times \boldvec{B}_0
\right) \ \cdot
\label{eq_emomentum}
\end{equation}
In the above equations, $c$ is the speed of light, $q_{\rm e} = -e$ the charge of
the electron, and $m_{\rm e}$ the mass of the electron.
In principle, the electric current should include contributions from both the
electrons and the ions. However, at the high frequencies considered here ($\omega
\ggg \omega_{\rm e}, \, |\Omega_{\rm e}|$; see
Sect.~\ref{sec:mathderivation_circular}), only the electrons are sufficiently
mobile to respond to the wave and contribute to the current -- the much more
massive ions remain virtually motionless.

Following the standard procedure, we look for complex solutions, $\boldvec{\sf
E}$, $\boldvec{\sf B}$, and $\boldvec{\sf V}_{\rm e}$, that vary as $\displaystyle
e^{i (\omega\,t - \boldvec{k} \cdot \boldvec{r}) }$,
where $\omega$ is the angular frequency ($\omega > 0$), $\boldvec{k}$ the wave
vector, and $\boldvec{r}$ the vector position of the considered point.%%\footnote{
We may then replace $\frac{\displaystyle \partial}{\partial t}$ by $i \omega$ and
$\nabla$ by $-i \boldvec{k}$ in
Eqs.~(\ref{eq_maxwell_ampere})--(\ref{eq_emomentum}).
In the case of parallel propagation, $\boldvec{k} \parallel \boldvec{B}_0
\parallel \boldhat{e}_z$, and these equations become
\begin{equation}
\omega \ \boldvec{\sf E}
\ = \ - c \, k_z \ \boldhat{e}_z \times \boldvec{\sf B} 
\ + \ 4\pi \, i \, q_{\rm e} \, n_{\rm e} \, \boldvec{\sf V}_{\rm e} \ ,
\label{eq_maxwell_ampere_complex}
\end{equation}
\begin{equation}
\omega \ \boldvec{\sf B}
\ = \ c \, k_z \ \boldhat{e}_z \times \boldvec{\sf E} \ ,
\label{eq_maxwell_faraday_complex}
\end{equation}
\begin{equation}
\omega \ \boldvec{\sf V}_{\rm e}
\ = \ -i \, \frac{q_{\rm e}}{m_{\rm e}} \ \boldvec{\sf E} 
\ - \ i \, \Omega_{\rm e} \ \boldvec{\sf V}_{\rm e} \times \boldhat{e}_z \ ,
\label{eq_emomentum_complex}
\end{equation}
where $\Omega_{\rm e} = \frac{\displaystyle q_{\rm e} B_0}{\displaystyle m_{\rm e}
c}$ is the electron gyro-frequency ($\Omega_{\rm e} < 0$).

The set of equations
(\ref{eq_maxwell_ampere_complex})--(\ref{eq_emomentum_complex}) admits two types
of solutions: 
electrostatic solutions, for which $\boldvec{\sf E}, \, \boldvec{\sf V}_{\rm e}
\parallel \boldhat{e}_z$ and $\boldvec{\sf B} = 0$, 
and electromagnetic solutions, for which $\boldvec{\sf E}, \, \boldvec{\sf B}, \,
\boldvec{\sf V}_{\rm e} \perp \boldhat{e}_z$.
Here, we only consider the latter.
Taking the cross-product of Eq.~(\ref{eq_emomentum_complex}) with $\boldhat{e}_z$,
we then obtain an equation for $\boldvec{\sf V}_{\rm e} \times \boldhat{e}_z$,
which can be re-injected into Eq.~(\ref{eq_emomentum_complex}) to give
\begin{equation}
\left( \omega^2 - \Omega_{\rm e}^2 \right) \, \boldvec{\sf V}_{\rm e}
\ = \ -i \, \omega \ \frac{q_{\rm e}}{m_{\rm e}} \ \boldvec{\sf E} 
\ - \ \Omega_{\rm e} \ \frac{q_{\rm e}}{m_{\rm e}} \ \boldvec{\sf E} \times
\boldhat{e}_z \ \cdot
\label{eq_emomentum_complex_mod}
\end{equation}
We now use Eqs.~(\ref{eq_maxwell_faraday_complex}) and
(\ref{eq_emomentum_complex_mod}) to eliminate $\boldvec{\sf B}$ and $\boldvec{\sf
V}_{\rm e}$ from Eq.~(\ref{eq_maxwell_ampere_complex}), whereupon we obtain an
equation for $\boldvec{\sf E}$ alone:
\begin{equation}
\omega^2 \ \boldvec{\sf E}
\ = \ c^2 \, k^2 \ \boldvec{\sf E} 
\ + \ \frac{\omega_{\rm e}^2 \ \omega^2}{\omega^2 - \Omega_{\rm e}^2} \ 
\left( 
\boldvec{\sf E} - i \, \frac{\Omega_{\rm e}}{\omega} \ \boldvec{\sf E} \times
\boldhat{e}_z 
\right) \ ,
\label{eq_maxwell_ampere_complex_mod}
\end{equation}
where $k = |k_z|$ is the wavenumber and $\omega_{\rm e}^2 =
\frac{\displaystyle 4\pi n_{\rm e} q_{\rm e}^2}{\displaystyle m_{\rm e}}$ the
plasma frequency squared.
Finally, in the same way as we proceeded with Eq.~(\ref{eq_emomentum_complex}),
we take the cross-product of Eq.~(\ref{eq_maxwell_ampere_complex_mod}) with
$\boldhat{e}_z$ 
and re-inject the resulting equation for $\boldvec{\sf E} \times \boldhat{e}_z$
into Eq.~(\ref{eq_maxwell_ampere_complex_mod}):
\begin{equation}
\left(
\omega^2 - c^2 \, k^2 - \frac{\omega_{\rm e}^2 \ \omega^2}{\omega^2 - \Omega_{\rm
e}^2} 
\right)^2 \, \boldvec{\sf E}
\ = \
\left(
\frac{\omega_{\rm e}^2 \ \Omega_{\rm e} \ \omega}{\omega^2 - \Omega_{\rm e}^2}
\right)^2 \, \boldvec{\sf E} \ \cdot
\label{eq_maxwell_ampere_complex_final}
\end{equation}

The dispersion relation directly follows from
Eq.~(\ref{eq_maxwell_ampere_complex_final}):
\begin{equation}
\omega^2 - c^2 \, k^2 - \frac{\omega_{\rm e}^2 \ \omega^2}{\omega^2 - \Omega_{\rm
e}^2} 
\ = \
\pm \ \frac{\omega_{\rm e}^2 \ |\Omega_{\rm e}| \ \omega}{\omega^2 - \Omega_{\rm
e}^2}
\label{eq_DR1}
\end{equation}
or, equivalently,
\begin{equation}
\omega^2 \ = \ c^2 \, k^2
\ + \ \frac{\omega_{\rm e}^2 \ \omega}{\omega \mp |\Omega_{\rm e}|} \ ,
\label{eq_DR2}
\end{equation}
which is identical to Eq.~(\ref{eq_DR}).
The $\mp$ sign in Eq.~(\ref{eq_DR2}) indicates the existence of two distinct
modes, the nature of which can be better understood by inserting
Eq.~(\ref{eq_DR1}) back into Eq.~(\ref{eq_maxwell_ampere_complex_mod}).
The result is simply
\begin{equation}
\boldvec{\sf E} = \pm \ i \, \boldvec{\sf E} \times \boldhat{e}_z \ ,
\label{eq_Efield_complex}
\end{equation}
which can be successively rewritten as
\begin{eqnarray}
\boldvec{\sf E}
& = &
{\sf E}_x \, \left( \boldhat{e}_x \mp i \, \boldhat{e}_y \right)
\nonumber \\
& = &
{\sf E}_{x0} \ e^{i (\omega\,t - k_z\,z)} \, 
\left( \boldhat{e}_x \mp i \, \boldhat{e}_y \right)
\nonumber \\
& = &
{\sf E}_{x0} \ \left[ e^{i \omega\,t} \ \boldhat{e}_x
+ e^{i ( \omega\,t \mp \frac{\pi}{2} )} \ \boldhat{e}_y \right] \
e^{-i k_z\,z} \ ,
\label{eq_Efield_complex_bis}
\end{eqnarray}
with ${\sf E}_{x0}$ generally complex.
Eq.~(\ref{eq_Efield_complex_bis}) clearly shows that $|{\sf E}_x| = |{\sf E}_y|$,
while ${\sf E}_y$ is behind [ahead of] ${\sf E}_x$ by a quarter period for the
mode corresponding to the upper [lower] sign.

Going back to physical space, the electric field vector, $\boldvec{E}$, is simply
the real part of the complex electric field vector, $\boldvec{\sf E}$.
It then follows that $E_x$ and $E_y$ have the same amplitude, while $E_y$ is
behind [ahead of] $E_x$ by a quarter period.
This means that $\boldvec{E}$ rotates circularly in a right-handed [left-handed]
sense about $\boldhat{e}_z$ -- which, we recall, was taken to lie in the
$+\boldvec{B}_0$ direction.
In other words, the wave is circularly polarized in a right-handed [left-handed]
sense about $\boldvec{B}_0$.
Therefore, the mode corresponding to the upper [lower] sign is called the right
[left] circularly polarized mode.

The reason for the existence of two circularly polarized modes can be found in the
electron momentum equation, Eq.~(\ref{eq_emomentum_complex_mod}), where the two
terms in the right-hand side represent the electric force and the magnetic force,
respectively.
Using Eq.~(\ref{eq_Efield_complex}), one can easily show that the latter is equal
to $\pm \frac{\displaystyle |\Omega_{\rm e}|}{\displaystyle \omega}$ times the
former.
Hence, the magnetic force is much smaller than the electric force, and it is
directed in the same sense as [the opposite sense to] the electric force for the
right [left] circularly polarized mode.
Physically, this is because the (dominant) electric force accelerates the
electrons in a right-handed [left-handed] sense about $\boldvec{B}_0$ for the
right [left] mode, whereas the magnetic force always acts on the electrons in a
right-handed sense about $\boldvec{B}_0$.

Had we taken the complex fields, $\boldvec{\sf E}$, $\boldvec{\sf B}$, and
$\boldvec{\sf V}_{\rm e}$, to vary as $\displaystyle e^{i (\boldvec{k} \cdot
\boldvec{r} - \omega\,t) }$, instead of $\displaystyle e^{i (\omega\,t -
\boldvec{k} \cdot \boldvec{r}) }$, we would have obtained the exact same
dispersion relation, Eq.~(\ref{eq_DR2}), but we would have found the opposite sign
in the right-hand side of Eq.~(\ref{eq_Efield_complex}), so that the last equality
in Eq.~(\ref{eq_Efield_complex_bis}) would have been
\begin{equation}
\boldvec{\sf E} \ = \
{\sf E}_{x0} \ \left[ e^{-i \omega\,t} \ \boldhat{e}_x
+ e^{-i ( \omega\,t \mp \frac{\pi}{2} )} \ \boldhat{e}_y \right] \
e^{i k_z\,z} \ \cdot
\nonumber
\end{equation}
In that case, it would have remained true that $E_y$ is behind [ahead of] $E_x$ by
a quarter period, and therefore that $\boldvec{E}$ rotates circularly in a
right-handed [left-handed] sense about $\boldvec{B}_0$, for the mode corresponding
to the upper [lower] sign.

\section{Electric field vector of a linearly polarized electromagnetic wave}
\label{sec:Efield_expression}

Consider again a cold, magnetized plasma with magnetic field $\boldvec{B}_0 = B_0
\, \boldhat{e}_z$, and now examine the parallel propagation ($\boldvec{k}
\parallel \boldvec{B}_0$) of a linearly polarized electromagnetic wave with
angular frequency $\omega$.
%%The electric field vector, $\boldvec{E}_\ell$, oscillates in a direction
$\boldhat{e}_\ell \perp \boldhat{e}_k$.
To fix ideas, assume that the electric field vector, $\boldvec{E}_\ell$, at the
source (subscript $\star$) oscillates along the $x$-axis with an amplitude $E_0$
and reaches a crest at the initial time, $t=0$:
\begin{equation}
\boldvec{E}_{\ell \star} = E_0 \ \cos (\omega\,t) \ \boldhat{e}_x \ \cdot
\label{eqB_Efield_lin_source}
\end{equation}
%%Eq.~(\ref{eqB_Efield_lin_source}) can be rewritten as
Let us now decompose the linearly polarized wave into a right (R) and a left (L)
circularly polarized mode:
\begin{equation}
\boldvec{E}_{\ell \star} = \boldvec{E}_{\rm R \star} + \boldvec{E}_{\rm L \star} \
\cdot
\label{eqB_Efield_lin_sum_source}
\end{equation}
%%with
By definition, $\boldvec{E}_{\rm R \star}$ [$\boldvec{E}_{\rm L \star}$] rotates
in a right-handed [left-handed] sense about $\boldvec{B}_0$.
Since $\boldvec{B}_0$ is in the positive $\boldhat{e}_z$ direction, we can write
\begin{equation}
\boldvec{E}_{\rm R \star} \ = \ {\textstyle \frac{1}{2}} \, E_0 \
\left[ \cos (\omega\,t) \ \boldhat{e}_x + \sin (\omega\,t) \ \boldhat{e}_y \right]
\label{eqB_Efield_R_source}
\end{equation}
and
\begin{equation}
\boldvec{E}_{\rm L \star} \ = \ {\textstyle \frac{1}{2}} \, E_0 \
\left[ \cos (\omega\,t) \ \boldhat{e}_x - \sin (\omega\,t) \ \boldhat{e}_y \right]
\ \cdot
\label{eqB_Efield_L_source}
\end{equation}

At a distance $s$ from the source, the electric field vector of the right [left]
mode, $\boldvec{E}_{\rm R}$ [$\boldvec{E}_{\rm L}$], is given by
Eq.~(\ref{eqB_Efield_R_source}) [Eq.~(\ref{eqB_Efield_L_source})], with time, $t$,
replaced by the retarded time, $\Big( t - {\displaystyle \int_0^s}
\frac{\displaystyle ds'}{\displaystyle V_{\phi,{\rm R}}} \Big)$ \Big[$\Big( t -
{\displaystyle \int_0^s} \frac{\displaystyle ds'}{\displaystyle V_{\phi,{\rm L}}}
\Big)$\Big], where $V_\phi = \frac{\displaystyle \omega}{\displaystyle k}$ is the
phase velocity.
This substitution is equivalent to replacing $(\omega\,t)$ by the phase,
$\phi_{\rm R}$ [$\phi_{\rm L}$], defined through Eq.~(\ref{eq_phase}), so that
\begin{equation}
\boldvec{E}_{\rm R} \ = \ {\textstyle \frac{1}{2}} \, E_0 \
\left[ \cos \phi_{\rm R} \ \boldhat{e}_x + \sin \phi_{\rm R} \ \boldhat{e}_y
\right]
\label{eqB_Efield_R}
\end{equation}
and
\begin{equation}
\boldvec{E}_{\rm L} \ = \ {\textstyle \frac{1}{2}} \, E_0 \
\left[ \cos \phi_{\rm L} \ \boldhat{e}_x - \sin \phi_{\rm L} \ \boldhat{e}_y
\right] \ \cdot
\label{eqB_Efield_L}
\end{equation}

The vector sum of Eqs.~(\ref{eqB_Efield_R}) and (\ref{eqB_Efield_L}) yields
\begin{eqnarray}
\boldvec{E}_\ell
& = & \boldvec{E}_{\rm R} + \boldvec{E}_{\rm L}
\nonumber \\
\noalign{\smallskip}
& = &
{\textstyle \frac{1}{2}} \, E_0 \
\left[
\left( \cos \phi_{\rm R} + \cos \phi_{\rm L} \right) \, \boldhat{e}_x
+ \left( \sin \phi_{\rm R} - \sin \phi_{\rm L} \right) \, \boldhat{e}_y
\right]
\nonumber \\
\noalign{\smallskip}
& = &
E_0 \ \cos \bar{\phi} \
\left( \cos {\textstyle \frac{\Delta \phi}{2}} \ \boldhat{e}_x + \sin {\textstyle
\frac{\Delta \phi}{2}} \ \boldhat{e}_y \right) \ ,
\label{eqB_Efield_lin_sum}
\end{eqnarray}
where $\bar{\phi} = \frac{\displaystyle \phi_{\rm R} + \phi_{\rm L}}{\displaystyle
2}$ is the average phase between the right and left modes
and $\Delta \phi = \phi_{\rm R} - \phi_{\rm L}$ is the phase difference between
them.
%%and $\Delta \psi^{\scriptscriptstyle [B]} = \frac{\displaystyle \phi_{\rm R} -
%%\phi_{\rm L}}{\displaystyle 2}$ is half the phase difference between them.
Hence, the right and left modes still superpose to create a linearly polarized
wave, but the polarization orientation has rotated by an angle $\frac{\Delta
\phi}{2}$ from the polarization orientation at the source (taken here to be along
$\boldhat{e}_x$; see Eq.~\ref{eqB_Efield_lin_source}).
This is exactly what we found in Sect.~\ref{sec:mathderivation_linear}, where we
also derived the expression of the Faraday rotation angle, $\frac{\Delta \phi}{2}$
(Eq.~\ref{eq_delta_psiB}).

For completeness, we now derive the expression of the phase $\bar{\phi}$ of the
linearly polarized wave.
From Eq.~(\ref{eq_phase}), it follows that
\begin{equation}
\bar{\phi} \ = \ \omega\,t - \int_0^s \bar{k} \, ds' \ ,
\label{eqB_phase_average}
\end{equation}
where $\bar{k} = \frac{\displaystyle k_{\rm R} + k_{\rm L}}{\displaystyle 2}$ is
the average wavenumber between the right and left modes.
The latter can be inferred from Eq.~(\ref{eq_phase_vel}) rewritten as an equation
for the wavenumber:
\begin{equation}
k \ = \ \frac{\omega}{c} \
\left( 1 - \frac{\omega_{\rm e}^2}{2 \, \omega^2}
\mp \frac{\omega_{\rm e}^2 \, |\Omega_{\rm e}|}{2 \, \omega^3}
\right) \ ,
\label{eqB_wavenumber}
\end{equation}
where the upper [lower] sign in the right-hand side pertains to the right [left]
mode:
\begin{equation}
\bar{k} \ = \ \frac{\omega}{c} \
\left( 1 - \frac{\omega_{\rm e}^2}{2 \, \omega^2}
\right) \ \cdot
\label{eqB_wavenumber_average}
\end{equation}
Thus, the wavenumber $\bar{k}$ and the phase $\bar{\phi}$ of the linearly
polarized wave are just the wavenumber and the phase of an electromagnetic wave
with the same $\omega$ in an unmagnetized plasma ($\Omega_{\rm e} =0$).

%%%%%%%%%%%%%%%%%%%%%%%%%%%%%%%%%%%%%%%%%%%%%%%%%%

% Don't change these lines
\bsp	% typesetting comment
\label{lastpage}
\end{document}